\documentstyle[12pt,psfig,oldlfont]{article}
\newcommand{\lsim}{\raisebox{-0.13cm}{~\shortstack{$<$ \\[-0.07cm] $\sim$}}~}
\newcommand{\gsim}{\raisebox{-0.13cm}{~\shortstack{$>$ \\[-0.07cm] $\sim$}}~}

\newcommand{\dx}{\mbox{\rm d}}
\newcommand{\ra}{\rightarrow}

\newcommand{\s}{\\ \vspace*{-3mm} }
\newcommand{\nn}{\noindent}
\newcommand{\non}{\nonumber}
\newcommand{\beq}{\begin{eqnarray}}
\newcommand{\eeq}{\end{eqnarray}}
\newcommand{\SM}{\mbox{${\cal SM}$}}
\newcommand{\SUSY}{\mbox{${\cal SUSY}~$}}
\newcommand{\MSSM}{\mbox{${\cal MSSM}$}}

\newcommand{\tg}{\mbox{tg}}
\newcommand{\tb}{\mbox{tg}\beta}
\newcommand{\ctb}{\mbox{ctg}\beta}

\newcommand{\ctg}{\mbox{ctg}}

\begin{document}

\begin{titlepage}

\begin{flushright}
DESY 95--211\\
KA--TP--9--95\\
IFT--95--14\\
November 1995 \\
\end{flushright}

\def\thefootnote{\fnsymbol{footnote}}

\vspace{1cm}

\begin{center}

{\large\sc {\bf  Two-- and Three--Body Decay Modes}}

\vspace*{3mm}

{\large\sc {\bf of SUSY Higgs Particles}}

\vspace{1cm}

{\sc A.~Djouadi$^{1,2}$\footnote{Supported by Deutsche Forschungsgemeinschaft
DFG (Bonn).}, J. Kalinowski$^{3}$\footnote{Supported in part by a KBN grant.}
and P.M.~Zerwas$^2$ }

\vspace{1cm}

$^1$ Institut f\"ur Theoretische Physik, Universit\"at Karlsruhe, \\
D--76128 Karlsruhe, FRG. \\
\vspace{0.3cm}

$^2$ Deutsches Elektronen--Synchrotron DESY, D--22603 Hamburg, FRG. \\
\vspace{0.3cm}

$^3$ Institute of Theoretical Physics,  Warsaw University, PL--00681 Warsaw,
Poland.

\end{center}

\vspace{1.6cm}

\begin{abstract}
\normalsize
\noindent

\nn We summarize the dominant decay modes of the neutral and charged Higgs
bosons in the Minimal  Supersymmetric extension of the Standard  Model. While
two--body decays are in general dominating, the branching ratios for
three--body decays of the  heavy  scalar, pseudoscalar and charged Higgs bosons
can be large below the thresholds if top quarks, $W/Z$ bosons or heavy scalar
bosons are involved. Analytical expressions have been derived for the partial
decay widths and the physical implications of these decay modes are discussed.

\end{abstract}

\end{titlepage}

\def\thefootnote{\arabic{footnote}}
\setcounter{footnote}{0}
\setcounter{page}{2}

\subsection*{1. Introduction}

The experimental exploration of the electroweak gauge symmetry breaking is one
of the most important  tasks in particle physics. The observation of one or
several fundamental scalar particles with couplings growing with the masses of
the sources would establish the Higgs mechanism as the physical basis of the
symmetry breaking. The Higgs mechanism \cite{R0} may be realized in the frame
of the Standard Model (\SM) or one of its possible extensions among which
supersymmetric theories are truly outstanding candidates
\cite{R1,R2}. To explore the physical nature of the scalar particles,
high-precision measurements of their properties are mandatory. To this end
precise calculations of the production cross sections and the branching ratios
of all important decay channels are required \cite{R2A}. \s

Since the Higgs  couplings to the Standard Model particles are proportional to
their masses [modulo enhancement/suppression factors in extended models] the
most important decay channels are two--body decays to the heaviest particles
allowed by phase space. Besides these two--body decays, below--threshold
three--body decays of the Higgs bosons can also be very important. This is
well-known in the Standard Model for Higgs decays to real and virtual  pairs of
$W/Z$ bosons in the intermediate mass range \cite{R8}: although suppressed by
the off--shell propagator and the additional electroweak coupling between the
$W/Z$ and the fermions, these decay processes are enhanced by the large Higgs
couplings to the gauge bosons, giving rise to appreciable branching ratios.\s

In supersymmetric extensions of the Standard Model, below--threshold decays may
become even more important, in particular for heavy Higgs decays to virtual and
real gauge boson pairs, mixed gauge and Higgs boson pairs, as well as top
quarks.  The  analysis of  three--body decays of the Higgs bosons in the
Minimal Supersymmetric Standard Model (\MSSM) has received not much  attention
in the literature so far. Only recently some below--threshold supersymmetric
Higgs boson decays have been investigated numerically in Ref.\cite{R9}. We
improve on this paper in several aspects: by performing the analysis
analytically; by completing the ensemble of important decay channels; and last
but not least, by studying the effects of stop mixing due to non--zero \SUSY
parameters $\mu$ and $A_t$. We will perform the analysis in the small and the
large  limit of $\tb$, the ratio of the vacuum expectation values, where the
results for three--body decays turn out to be quite different. Both domains are
interesting since they are realized in grand unified supersymmetric models
with $b$--$\tau$ Yukawa coupling unification \cite{R6A}.\s

The paper is organized as follows. In the next Section, we summarize the main
features of the Higgs sector in the \MSSM . In Section 3 we discuss the
branching ratios in the large $\tb$ case, with special emphasis on decays for
Higgs masses at the edge of the \SUSY parameter space. In Section 4 we analyze
the below-threshold decays of the heavy CP-even, CP-odd and charged Higgs
bosons for small $\tb$ values. The total decay widths are summarized in Section
5. Supplementing analytical expressions will be presented in the Appendix.

\subsection*{2. Physical Set--Up}

In the Minimal Supersymmetric extension of the Standard Model two isodoublets
of Higgs fields \cite{R1} are introduced to provide masses to the up-- and
down--type fermions. This results in a spectrum of a quintet of physical
particles: two CP--even neutral scalars $h$ and $H$, one CP--odd neutral
(pseudo)scalar $A$, and a  pair of charged scalar particles $H^\pm$. Besides
the four masses, two additional parameters determine the properties of these
particles at the tree level: the ratio $\tb$ of the vacuum expectation values
of the two neutral Higgs fields and a mixing angle $\alpha$ in the neutral
CP--even sector. \s

Supersymmetry leads to several relations among the parameters of the  Higgs
sector, and only two of them are in fact independent. If one of the Higgs boson
masses [in general  $M_A$] and $\tb$ are specified, all other masses and the
mixing angle  $\alpha$ can be derived at the tree--level \cite{R2}.
Supersymmetry imposes a strong  hierarchical structure on the mass spectrum,
$M_h<M_A<M_H$, $M_W< M_{H^\pm}$ and $M_h<M_Z$, which however is broken by
radiative corrections \cite{R3,R4}. \s

The leading part of the  radiative corrections grows as the fourth power of
the top quark mass $m_t$ and the logarithm of the squark mass $M_S$
\cite{R3}. This part is determined by the parameter
\begin{eqnarray}
\epsilon = \frac{3 G_F}{\sqrt{2} \pi^2} \frac{m_t^4}{\sin^2 \beta} \log
\left( 1+ \frac{M_S^2}{m_t^2} \right)
\end{eqnarray}
These leading corrections can be summarized in a simple form \cite{R3}
\begin{eqnarray}
M^2_{h} & = & \frac{1}{2} \left[ M_{A}^2 + M_Z^2 + \epsilon \right.
\non \\
& & \left. - \sqrt{(M_{A}^2+M_Z^2+\epsilon)^2
-4 M_{A}^2 M_Z^2 \cos^2 2\beta
-4\epsilon (M_{A}^2 \sin^2\beta + M_Z^2 \cos^2\beta)} \right] \non \\
M_{H}^2 & = & M_{A}^2 + M_Z^2 - M_{h}^2 + \epsilon \non \\
M_{H^\pm}^2 & = & M_{A}^2 + M_W^2 \\
\tg 2 \alpha &=& \tg 2\beta \frac{M_{A}^2 + M_Z^2}{M_{A}^2 - M_Z^2 +
\epsilon/\cos 2\beta} \ \  \ \  \ \ \  \left[ -\frac{\pi}{2} < \alpha <0
\right]
\end{eqnarray}

At the subleading level the radiative corrections are affected by the
supersymmetric Higgs mass parameter $\mu$ and the parameter $A_t$ in the soft
symmetry breaking interaction\footnote{We adopt the recent analysis of
Ref.\cite{R5} where the full dependence on $\mu, A_t$ is taken into account
and where the next--to--leading QCD corrections \cite{R4} are implemented.}.
The radiative corrections are positive and they shift the mass of the light
neutral Higgs boson $h$ upward. The variation of the Higgs boson masses $M_h$,
$M_H$, $M_{H^\pm}$ with the pseudoscalar mass $M_A$ is shown in Fig.~1 for $\tb
=1.5$ and 30 in two mixing scenarios \cite{R5}: (i) (practically\footnote{A
value of $\mu \sim 100$ GeV is not experimentally excluded yet; the partial
decay widths we are analyzing, change only slightly between $\mu=0$ and 100
GeV.}) no mixing $\mu \ll M_S$, $A_t=0$; and (ii) maximal mixing $\mu \ll
M_S$,~$A_t=\sqrt{6} M_S$. The top quark mass is fixed to $m_t=175$ GeV and the
scalar mass to  $M_S=$ 1 TeV. If $M_A$ is large, the $A$, $H$, $H^{\pm}$ Higgs
bosons are nearly degenerate while the lightest Higgs boson $h$ reaches its
maximal mass value. Note that for small $\tb$ the ordering
$M_H>M_{H^{\pm}}>M_A$ holds in the large mass range while $M_{H^{\pm}}>M_H
\simeq M_A$ for large $\tb$ instead. The effect of non-zero $A_t$ and $\mu$ is
quite significant since its shifts the maximal value of the $h$ mass upward
by almost $\sim 20$ GeV; for large $\tb$ values, the relations $M_h \simeq
M_A$, $M_H \simeq$ max$(M_h)$ for $M_A <$ max$(M_h)$ and $M_H \simeq M_A$,
$M_h \simeq $ max$(M_h)$ for $M_A >$ max$(M_h)$, holding in the case of no
mixing, still hold true when the mixing is included. \s

The \underline{couplings of the neutral Higgs bosons to fermions and gauge
bosons} depend on the  angles $\alpha$ and $\beta$; they are given in Table 1
with the normalization defined by the  \SM\ couplings
\begin{eqnarray}
g_{H_{{\cal SM}}ff} = \left[ \sqrt{2} G_F \right]^{1/2} m_f \ \ \  \ \ \
\mbox{\rm and} \ \ \
g_{H_{{\cal SM}}VV} = 2 \left[ \sqrt{2} G_F \right]^{1/2} M_V^2
\end{eqnarray}
The CP--even neutral  Higgs bosons $h$, $H$  share the \SM\ coupling to the
massive gauge bosons, their couplings to down-- (up--) type fermions are
enhanced  (suppressed) with respect to the \SM\ case. If, in the case of large
$A$ masses, the mass of the lightest Higgs boson is  close to its upper limit
for a given value of $\tb$, the   couplings of $h$ to fermions and gauge bosons
are \SM --like  while the couplings of the heavy CP--even scalar $H$ are
suppressed. The pseudoscalar Higgs boson  has no tree-level couplings to gauge
bosons; its couplings to down-- (up)--type fermions are (inversely)
proportional to  $\tb$. Radiative corrections treated at the level discussed
above, are incorporated in the mixing angle $\alpha$. The size  of the
couplings is shown in Fig.~2a for the mixing scenarios defined before. Here
again, the mixing has a large impact; however the bulk of the effect
consists of shifting the $h,H$ masses upward.

\vspace*{3mm}

\begin{center}
\begin{tabular}{|c||c|c||c|c|} \hline
& & & \s
$\hspace{1cm} \Phi \hspace{1cm} $ &$ g_{ \Phi \bar{u} u} $ & $
g_{\Phi \bar{d} d} $ & $g_{ \Phi VV} $ \\
& & & \\ \hline \hline
& & & \\
$H_{{\cal SM}}$ & \ $ \; 1  \;     $ \ & \ $ \; 1  \; $ \  & \ $ \; 1  \;
 $ \ \\[0.3cm] \hline
& & & \\
$h$  & \ $\; \cos\alpha/\sin\beta       \; $ \ & \ $ \; -\sin\alpha/
\cos\beta \; $ \ & \ $ \; \sin(\beta-\alpha) \; $ \ \\
 $H$  & \       $\; \sin\alpha/\sin\beta \; $ \ & \ $ \; \cos\alpha/
\cos\beta \; $ \ & \ $ \; \cos(\beta-\alpha) \; $ \ \\
$A$  & \ $\; 1/ \tb \; $        \ & \ $ \; \tb \; $ \
& \ $ \; 0 \; $ \ \\[0.3cm] \hline
\end{tabular}
\end{center}

\vspace*{3mm}

\nn {\small Tab.~1: {\it
Higgs boson couplings in the \MSSM\ to fermions and gauge
bosons relative to the ${\cal SM}$ Higgs  couplings.}}

\vspace*{5mm}

In the limit of large $A,H$ masses, the $h$ mass becomes maximal. In this
limit, the properties of $h$ approach those of the \SM~Higgs boson. The heavy
neutral scalar Higgs boson $H$ decouples from the gauge bosons and the
fermionic $H$ couplings approach the corresponding couplings of the
pseudoscalar
Higgs boson $A$. For small $h,H$ masses, the light Higgs boson $h$ is built--up
primarily by the Higgs field $H_2$ which couples to down--type fermions with
the
strength $ \sim \tb$, while the heavy Higgs boson $H$, built--up by $H_1$,
couples conversely to up--type fermions. \s

For fermions the \underline{charged  Higgs  particles} couple to  mixtures  of
scalar and pseudoscalar currents, with  components proportional to $m_u\ctb$
and $m_d\tb$ for the two $\pm$ chiralities,
\begin{eqnarray}
g_{H^+u\bar{d} } = \left( G_F/\sqrt{2} \right)^{1/2} \left[(1-\gamma_5)
m_u \ctb + (1+\gamma_5) m_d \tb \right]
\end{eqnarray}

The \underline{couplings of two Higgs bosons with one gauge boson} are
listed in Table 2. They are normalized to the charged/neutral weak couplings
\beq
\label{eq1}
g_W=( \sqrt{2} G_F)^{1/2}M_W \ \ \ \mbox{and} \ \ \ g_Z=(\sqrt{2}G_F)^{1/2}M_Z
\eeq
and they come with the sum of the Higgs momenta entering and leaving the
vertices. Again, radiative corrections are incorporated
in the mixing angle $\alpha$. The
magnitude of these couplings can be read off Fig.~2a.

\vspace*{3mm}

\begin{center}
\begin{tabular}{|c||c|c|} \hline
& & \\
$\hspace{1cm} \Phi \hspace{1cm}$ & $ g_{\Phi AZ}/g_Z $ & $g_{\Phi H^\pm W^\pm}
/g_W$ \\[0.3cm] \hline \hline & & \\
$h$ & \ $ \; \ \cos (\beta-\alpha)  \; $ \ & \ $ \;
\mp \cos (\beta-\alpha)  \; $ \\
$H$ & \ $ \; - \sin (\beta-\alpha) \; $ \ & \ $ \;
\pm \sin (\beta-\alpha)  \; $ \\
$A$ & \ $ \;   0 \; $ \ & \ $ \; 1  \; $ \\[0.3cm] \hline
\end{tabular}
\end{center}

\vspace*{3mm}

\nn {\small Tab.~2: {\it
The couplings of two Higgs bosons with one gauge boson.
They are normalized to the weak couplings defined in eq.(\ref{eq1}) and
they come with  momenta of the Higgs particles entering and leaving the
vertices.}}

\vspace*{5mm}

Finally, we summarize the couplings of the three neutral Higgs bosons among
themselves which we will need in the subsequent analyses. Unlike the previous
cases, the  radiative corrections  are not incorporated in the mixing angle
$\alpha$ in total, but additional contributions must be taken into account
explicitely. Normalized to
\beq
g_Z'=(\sqrt{2}G_F)^{1/2}M_Z^2
\eeq
the radiatively corrected self--couplings are   to leading order \cite{R6}
\begin{eqnarray}
\lambda_{Hhh} &=& 2 \sin 2\alpha \sin (\beta+\alpha) -\cos 2\alpha \cos(\beta
+ \alpha) + 3 \frac{\epsilon}{M_Z^2} \frac{\sin \alpha}{\sin\beta}
\cos^2\alpha \label{3h2}\\
\lambda_{HAA} &=& - \cos 2\beta \cos(\beta+ \alpha)+ \frac{\epsilon}{M_Z^2}
\frac{\sin \alpha}{\sin\beta} \cos^2\beta \label{3h3} \non
\end{eqnarray}
These couplings are shown in Fig.~2b for the two mixing scenarios, including
subleading contributions \cite{R5} for non--zero $\mu$ and $A_t$. \s

The value of $\tb$ determines to a large extent the decay pattern of the
supersymmetric Higgs bosons. For large values of $\tb$ the pattern is simple, a
result of the strong enhancement of the Higgs couplings  to down--type
fermions. The neutral Higgs bosons will decay into $b\bar{b}$ and $\tau^+
\tau^-$ pairs, the charged Higgs bosons into $\tau \nu_\tau$ pairs below and
$tb$ pairs above the top--bottom threshold. Only at the edges of the Higgs
parameter space these simple rules are modified: If $h$ approaches the maximal
mass value, the couplings become \SM --like and the decay modes follow the
pattern of the Standard Model; if $H$ approaches the minimal mass value, it
will mainly decay into $hh$ and $AA$ final states.
The detailed analysis of the large $\tb$ scenario with special emphasis on
\begin{eqnarray}
h  & \ra &  W^* W^*, Z^* Z^* \ra {\rm 4~fermions} \\
H  & \ra& hh^*, AA^* \ra hb\bar{b},Ab\bar{b}
\end{eqnarray}
is presented in Section 3. \s

For small values of $\tb\sim 1$ the decay pattern of the heavy neutral Higgs
bosons  is much more complicated \cite{R6,R7}. The $b$ decays are in general
not dominant any more. Instead, cascade decays to pairs of light Higgs
bosons and mixed pairs of Higgs and gauge bosons are important. Moreover,
decays to gauge boson pairs play a major role. However, for very
large masses, the
neutral Higgs bosons decay almost exclusively to top quark pairs. The decay
pattern of the charged Higgs bosons for small $\tb$ is similar to that at large
$\tb$ except in the intermediate mass range where cascade decays to light Higgs
and $W$ bosons are dominant for small $\tb$. \s

Besides these two--body decays, below--threshold three--body decays of Higgs
bosons can play an important role. ${\cal SM}$ Higgs decays into real and
virtual $Z$  pairs \cite {R8} provide the signature for searching these
particles in the intermediate mass range at proton colliders.
The suppression by
the off--shell propagator and the additional electroweak coupling between the
$Z$  boson and the  fermions is at least partly compensated by the large
Higgs coupling to the $Z$ bosons.
By the same token, three--body decays of \MSSM\ Higgs particles mediated
by gauge bosons, heavy Higgs bosons and top quarks, are of physical
interest, specifically, Fig.~3: \s

\nn (i) \underline{heavy  CP--even  Higgs particle $H$:}
\begin{eqnarray}
H \ &  \ra & \ VV^* \ \ra V f \bar{f}^{(')} \\
H \ &  \ra & \ AZ^* \ \ra  \ A f \bar{f} \\
H \ &  \ra & \ H^\pm W^{\mp *} \ \ra  \ H^\pm f \bar{f}' \\
H \ & \ra &  \ \bar{t}t^* \  \ra \ \bar{t} b W^+
\end{eqnarray}
(ii)  \underline{CP--odd Higgs boson $A$: }
\begin{eqnarray}
A \ & \ra & \ hZ^* \ \ra \ h f \bar{f} \\
A \ & \ra &  \ \bar{t}t^* \  \ra \ \bar{t} bW^+ \hspace*{2cm}
\end{eqnarray}
(iii) \underline{charged Higgs boson $H^{\pm}$:}
\begin{eqnarray}
\hspace*{1.5cm} H^\pm \ & \ra & \ hW^* \ \ra \ h f \bar{f}' \\
\hspace*{1.5cm} H^\pm \ & \ra & \ AW^* \ \ra \ A f \bar{f}' \\
\hspace*{1.5cm} H^\pm \ & \ra & \ \bar{b}t^* \ \ra \ \bar{b}bW  \hspace{4cm}
\end{eqnarray}
The decays $H\ra W^+\bar{t}b$ can also be mediated by virtual $W^+H^{-*}$
and $W^+W^{-*}$ intermediate states, and likewise
$A\ra W^+\bar{t}b$ decays by $W^+H^{-*}$
states. However, these contributions, given in
the Appendix, are very small. [Additional three--body decays, $H\ra hh^*$
or $AA^*$ with $h^*,A^* \ra b\bar{b}$, could  be relevant only for $M_H$ a few
hundred MeV from its minimum value if  $\tb$ were very large.]
Three--body decays of the light CP--even Higgs
boson $h$ are negligible, with the only exception of $h \ra W^{*} W^{*}$ for
$M_h$ close to its maximal value.  \s

The decay chains given in the listings above for the heavy Higgs bosons are the
dominant mechanisms in the range of Higgs masses where
multi--body decays are relevant.
This may be exemplified for the first decay chain of $A$.
In a microscopic analysis one
would consider the chain $A\ra h^*Z^*\ra (b\bar{b})_h (f\bar{f})_Z$. However,
for small values of $\tb$, the coupling $hbb$ is of order $\sqrt{G_F} m_b$
and much smaller than the gauge coupling $Zff$. As a result, the
off--shell $h^*$ contributions are negligible compared to the off--shell $Z^*$
contributions below the two--particle threshold. The light Higgs boson $h$ can
therefore be treated as a stable particle and the decay process can be
described as a three--body process to a high level of accuracy. This feature
enables us to approach the problem of below--threshold decays by simple
analytical methods. \s

In the discussion so far we have disregarded Higgs decays to neutralinos,
charginos and sfermions. Since we have assumed a common \SUSY
scale of order
$M_S \sim 1$ TeV, decays to sfermions do not play a role
in the Higgs mass range
of a few hundred GeV which we are analyzing.
If the chargino/neutralino channels are
kinematically open, they affect the decay branching ratios in general very
strongly \cite{R7}. Denoting the decay branching ratio without
neutralino/char\-gi\-no decays by $BR$ and the branching ratio for
neutra\-lino/char\-gino (and possibly sfermion) decays by $BR_{\chi}$,
the finally observed
branching ratio is given by
\beq
BR_{fin}=(1-BR_{\chi}) \times BR
\eeq
The branching ratios $BR_{\chi}$ depend on the supersymmetric Higgs parameter
$\mu$ and the gaugino mass parameter $M_2$. For large values of $\mu$ and $
M_2$, these decay modes can be neglected, yet for parameter sets $\mu\sim M_2
\sim {\cal O}(100)$ GeV, they could eventually dominate over all other decay
modes \cite{R7}. Since the aim of the present paper is to identify the regions
in the parameter space where the below--threshold decay modes might be
important, we will assume, in the subsequent discussion,
the decay channels  to supersymmetric particles to be shut. \s


\subsection*{3. Higgs decay branching ratios: Large $\tb$}

For large $\tb$ the Higgs couplings to down--type fermions dominate over all
other couplings. As a result, the decay pattern is in general very simple. The
neutral Higgs bosons will decay into $b\bar{b}$ and $\tau^+\tau^-$ pairs for
which the branching ratios are close to $\sim 90$~\% and $\sim 10$~\%,
respectively. In terms of the Fermi decay constant $G_F$, the partial decay
widths into fermions are given by \cite{R2}
\beq
\Gamma(\Phi \ra \bar{f}f) = N_c \frac{G_F M_\Phi }{4 \sqrt{2} \pi} g^2_{\Phi
ff}
m_f^2 \beta^p
\eeq
with $p=3(1)$ for the CP--even (odd) Higgs bosons;  $\beta=(1-4m_f^2/
M_\Phi^2) ^{1/2}$ is the velocity of the fermions in the final state, $N_c$ the
color factor. The couplings $g_{\Phi ff}$ are collected in Table 1. For the
decay widths into quark  pairs, the QCD radiative corrections \cite{R10a} are
large and must be included. In the limit $M_\Phi \gg m_q$, the ${\cal O}
(\alpha_s^2)$ corrected decay widths read \cite{R11}
\beq
\Gamma (\Phi \ra q \bar{q}) = \frac{3 G_F M_\Phi}{ 4\sqrt{2} \pi} m_q^2
g^2_{\Phi qq} \left[ 1+5.67 \left( \frac{\alpha_s}{\pi} \right)+
(35.94-1.36N_F) \left( \frac{\alpha_s}{\pi} \right)^2 \right]
\eeq
with $\alpha_s \equiv \alpha_s(M_\Phi)$, $m_q^2\equiv m_q^2(M_\Phi)$; $N_F=5$
is the number of active quark flavors; all quantities are defined in the
$\overline{{\rm MS}}$ scheme with $\Lambda^{(5)}_{\overline{{\rm MS}}}
\sim 226$ MeV
[which corresponds to the value $\alpha_s(M_Z^2)=0.118$]. The bulk
of these QCD corrections can be absorbed into running quark masses evaluated
at the scale $\mu=M_\Phi$:
\beq
m_q(M_\Phi)=m_q(m_q) \left[\frac{\alpha_s(M_\Phi)}{\alpha_s(m_q)}
\right]^{12/(33-2N_F)} \times \frac{ 1+ c_1^q\alpha_s(M_\Phi)/\pi
+c_2^q\alpha_s^2(M_\Phi)/\pi^2}{1+ c_1^q\alpha_s(m_q)/\pi+c_2^q \alpha_s^2
(m_q)/\pi^2}
\eeq
In the case of bottom (charm) quarks, the coefficients are
$c_1=1.17(1.01)$ and
$c_2=1.50(1.39)$. For $M_\Phi \simeq 100$ GeV, the $b$ and $c$ quark masses
$m_b(m_b)=4.23$ and $m_c(m_c)=1.23$ GeV, as extracted from QCD sum rules
\cite{R12}, have dropped to the effective values\footnote{A detailed
analysis of QCD effects in Higgs decays will be presented in
Ref.\cite{R13A}.} $m_b(M_\Phi) \simeq 2.9$ GeV
and $m_c (M_\Phi) \simeq 0.62$ GeV. \s

The charged Higgs particles decay into $\tau\nu_{\tau}$ pairs below and
into $tb$
pairs above the top--bottom threshold. [Virtual $t^*\ra bW$ contributions give
rise to three--body decays for charged Higgs decays $H^{\pm}\ra t^*b \ra
Wb\bar{b}$ only a few GeV below the two-body decay threshold for $tb$ final
states.] Neglecting one of the fermion masses in the phase space [$m_\nu=0$ and
$m_b \ll m_t$], the partial decay widths are \cite{R2}
\begin{eqnarray}
\Gamma(H^+ \rightarrow u \bar{d}) =  \frac{N_c G_F}{4 \sqrt{2} \pi} M_{H^\pm}
|V_{ud}|^2 \ \left( 1- \frac{m_f^2}{M_{H^\pm}^2} \right)^3 \left[ m_{d}^{2}
{\tg}^2\beta + m_u^2 {\ctg}^2 \beta  - 4\frac{m_u^2 m_d^2}{M_{H^\pm}^2} \right]
\end{eqnarray}
with $V_{ud}$ being the CKM--like matrix element introduced for quark final
states. Also in this case the QCD corrections \cite{R10b} must be included. For
light quarks, the bulk of these corrections can be taken into account by using
running quark masses; for the top quark, the QCD corrections are of order
$\alpha_s/\pi$ and small, since the top quark mass is of the same order as the
Higgs boson mass. \s

The branching ratios for these decays are shown in Fig.~4 for $\tb=30$.
Other decay modes become important only at the edges of the \SUSY
parameter
space: (i) if $h$ approaches the maximum mass for a given value of $\tb$, and
(ii) if $H$ approaches the minimum mass for a given value of
$\tb$. These two cases are
discussed below in detail.

\subsubsection*{3.1 The light neutral Higgs boson $h$}

If the lightest CP--even Higgs boson $h$ approaches its maximum mass for a
given value of $\tb$, the couplings become ${\SM}$--like
and the decay pattern of
the particles is the same as for the Higgs particle in the Standard Model. In
this part of the parameter space, besides decays into charm quarks and
gluons\footnote{The expression for the gluonic decay width, including the QCD
corrections, can be found in Ref.\cite{R13}.}, neutral Higgs decays to pairs of
real and virtual gauge bosons would be observed. Allowing both gauge bosons to
be off--shell [important only for $h$ masses
close to $M_W$ or $M_Z$], the partial decay width reads \cite{R14}
\beq
\Gamma(h \ra V^* V^*) = \frac{1}{\pi^2} \int_0^{M_h^2} \frac{ {\rm d}Q_1^2 M_V
\Gamma_V } {(Q_1^2-M_V^2)^2+M_V^2 \Gamma_V^2} \int_0^{(M_h-Q_1)^2} \frac{{\rm
d}
Q_2^2 M_V \Gamma_V } {(Q_2^2-M_V^2)^2 +M_V^2 \Gamma_V^2} \ \Gamma_0
\eeq
with $Q_1^2,Q_2^2$ being  the squared invariant masses of the
virtual gauge bosons, $M_V$ and
$\Gamma_V$ their masses and total decay widths. $\Gamma_0$ is given by
\beq
\Gamma_0 = \frac{G_F M_h^3} {16 \sqrt{2} \pi} \delta_V'\ \sin^2(\beta-\alpha)
\lambda^{1/2} (Q_1^2,Q^2_2;M_h^2) \left[ \lambda (Q_1^2,Q^2_2;M_h^2)
+ \frac{12Q_1^2 Q_2^2}{M_h^4} \right]
\eeq
with
$\delta_V'=2(1)$ for $V=W(Z)$; $\lambda$ is the two--body phase space
function defined by
\beq
\lambda(x,y;x)= (1-x/z-y/z)^2-4xy/z^2 \label{lamb}
\eeq

The branching ratios for the lightest CP--even Higgs boson
are shown as a function  of $M_h$ near its
maximal allowed mass value [in practice for $M_A=1$ TeV]  in Fig.~5a.
The biggest uncertainty in the prediction of the decay branching ratios is
associated with the badly known charm quark mass and the QCD coupling constant.
To exemplify\footnote{We refrain from discussing the migration of the
uncertainty in the charm sector to the other branching ratios which will be
displayed in the figure for the central values of the coupling and mass
parameters. A comprehensive discussion is presented in Ref.\cite{R13A}.}
the size of the uncertainty we have indicated the error of the
branching ratio for $h \ra c\bar{c}$ by the shaded band in Fig.~5a. The
parameters chosen in the figure are the $\overline{\rm MS}$ QCD coupling
$\alpha_s (M_Z^2)=0.118 \pm 0.006$ and $m_c= 1.23 \pm 0.03$ GeV for the
charm quark mass in the $\overline{\mbox{MS}}$ scheme at the
scale of the pole mass as extracted from
QCD sum rules \cite{R12}. The partial charm decay width is calculated
consistently to ${\cal O}(\alpha_s^2)$ with the renormalization scale set to
the Higgs boson mass \cite{R13A} (see also \cite{R15}).
It is evident from Fig.~5a that the branching
ratio for $h\ra c\bar{c}$ is very  uncertain. Since it is small, the migration
of the error to the $b$ and $\tau$ branching ratios is less important,
however.

\subsubsection*{3.2 The heavy neutral Higgs boson $H$}

 If the mass of the heavy neutral CP--even Higgs boson is close
to the minimal value [for $\tb \gg 1$, this corresponds to $M_A \simeq M_h
<{\cal O} (100$ GeV)], the
main decay modes are the cascade decays $H \ra \Phi
\Phi$ with $\Phi=h$ or $A$.
For real light Higgs bosons, the partial decay widths are given by \cite{R2}
\begin{eqnarray}
\Gamma(H \ra \Phi \Phi) = \frac{G_F}{16\sqrt{2} \pi} \frac{M_Z^4}{M_H}
\left(1-4\frac{M_\Phi^2}{M_H^2} \right)^{1/2} \lambda^2_{H\Phi \Phi}
\label{HAA}
\end{eqnarray}
where the radiatively corrected three--boson self--couplings have been given
to leading order in eqs.(\ref{3h2},\ref{3h3}) and including subleading
terms in Fig.~2b. Since $M_h \simeq M_A$ and $\lambda_{Hhh}
\simeq \lambda_{HAA}$ for large values of $\tb$,
the two branching ratios are equal and close to 50\% each. In this
range of small $H$ masses, the decays $H \ra \Phi \Phi^*$ with $\Phi^* \ra b
\bar{b}$, Fig.~3, can be important since the $\Phi bb $ coupling $\sim \tb
\gg 1$ is large. Taking into account only the diagrams where the
$b\bar{b}$ final states originate from the decay of $h^*$ or $A^*$ [the
additional channels where $h$ or $A$ are emitted from the $b$ lines in $H \ra
b\bar{b}$ give negligible contributions since the $Hbb$ coupling is
very small in this mass range, see Fig.~2a], the Dalitz plot density for this
three--body decay is given by
\beq
\frac{\dx \Gamma }{\dx  x_1 \dx x_2}(H \ra \Phi\Phi^*) = K_{H \Phi \Phi}
\frac{x_1+x_2-1+\kappa_{\Phi}}{(1-x_1-x_2)^2 +\kappa_\Phi \gamma_\Phi}
\eeq
where $x_{1,2}=2E_{1,2}/M_H$ are the scaled energies of the fermions
 in the final state [which we take to be massless],
and $\kappa_{\Phi}=M^2_{\Phi}/M^2_H, \gamma_\Phi=\Gamma_\Phi^2/M_H^2$
for $\Phi=h,A$. The coefficient $K_{H \Phi \Phi}$ is given by
\beq
K_{H \Phi \Phi} = \frac{3G_F^2} {16\pi^3} \ \frac{M_Z^4}{M_H} \
g^2_{\Phi bb} m_b^2 \ \lambda^2_{H\Phi\Phi}
\eeq
The integration of the Dalitz density over the energies $x_1$ and $x_2$
between the boundaries
\begin{eqnarray}
1- x_2 -\kappa_{\Phi} < &x_1& < 1- \frac{\kappa_{\Phi}}{1-x_2} \non \\
0 < &x_2& < 1- \kappa_{\Phi}  \label{boundary}
\end{eqnarray}
can be performed analytically, with the result for $M_H \lsim
2 M_\Phi-\Gamma_\Phi$
sufficiently below the thresholds
\beq
\Gamma(H\ra \Phi \Phi^*) &=& K_{H \Phi \Phi} \left[ (\kappa_\Phi -1) \left( 2-
\frac{1}{2} \log \kappa_\Phi \right) + \right. \non \\
&&\left. \frac{1-5 \kappa_\Phi}{\sqrt{4\kappa_\Phi -1}} \left( \arctan
 \frac{2 \kappa_\Phi -1} {\sqrt{4\kappa_\Phi
-1}}-\arctan\frac{1}{\sqrt{4\kappa_\Phi -1}} \right) \right]
\eeq
The branching ratios for these two decay modes are shown in Fig.~5b for
$\tb=30$ as a function of the pseudoscalar mass and for the two mixing
scenarios discussed previously.

\subsection*{4. Higgs decay branching ratios: Small $\tb$}

In the case  $\tb \gsim 1$ the Higgs decay pattern is more complicated because
the couplings to the light $b$ quarks and $\tau$ leptons are not enhanced any
more and the couplings to the heavy top quark and the massive gauge bosons
become increasingly important, leading in a natural way to a significant
admixture of three--body decays.

\subsubsection*{4.1 The CP--even Higgs bosons}

Even for small values of $\tb$, $\tb \gsim 1$, the \underline{light neutral
CP-even Higgs boson $h$} decays primarily to $b$ quark and $\tau$ pairs. Only
near its maximum mass limit, where the properties of $h$ are ${\cal SM}$--like,
other decay modes become non--negligible, as discussed before. The branching
ratios are shown in Fig.~6a for $\tb=1.5$ in the cases of ``no--mixing" and
``maximal mixing" introduced above. \s

For the \underline{heavy neutral CP--even Higgs boson $H$}, besides the usual
2--fermion $b\bar{b}$, $c\bar{c}$, $\tau\tau$ and the $gg$ decays, important
channels are decays to pairs of light Higgs and gauge bosons and mixed pairs of
Higgs and gauge bosons. Above the $t\bar{t}$ threshold, the Higgs boson $H$
decays almost exclusively to top quarks. We will concentrate on
below--threshold 3--body decays involving gauge bosons, Higgs bosons and top
quarks, Fig.~3.

\bigskip

\nn {\it (i) Gauge boson pairs}

\bigskip

In the mass range above the $WW$ and $ZZ$ thresholds, where   the $HVV$
couplings are not strongly suppressed for small values of $\tb$, the partial
widths of the heavy Higgs particle $H$ for decays into massive gauge
bosons may be written
\begin{eqnarray}
\Gamma (H \ra VV) =  \frac{\sqrt{2}G_F M_H^3}{32 \pi}
\cos^2 (\alpha-\beta) (1-4\kappa_V+12\kappa_V^2)
(1-4\kappa_V)^{1/2} \ \delta_V'
\end{eqnarray}
with $\kappa_V=M_V^2/M_H^2$ and $\delta_V'=2(1)$ for  $V=W(Z)$.
Below the threshold for two real bosons, the decays to real and virtual
$VV^*$ pairs are important down to Higgs masses $M_H\sim 130 $ GeV.
The Dalitz plot density for the decay process $H\ra VV^*\ra Vf\bar{f}^{(')}$
is given by
\begin{eqnarray}
\frac{ \dx \Gamma }{\dx  x_1 \dx x_2} (H \ra V V^*) = K_{HVV} \
\frac{(1-x_1)(1-x_2) +\kappa_V (2x_1+2x_2-3+2 \kappa_V)}{(1-x_1-x_2)^2
+ \kappa_V \gamma_V}
\end{eqnarray}
for all final $f\bar{f}'$ fermions taken massless [$i.e.$ $t$ quark excluded]
and summed up.
The overall normalization factor $K_{HVV}$ is defined as
\begin{eqnarray}
K_{HVV} = \frac{3 G_F^2 M_W^4}{16\pi^3 } \cos^2(\beta- \alpha) M_H \delta_V
\end{eqnarray}
with
\begin{eqnarray}
\delta_W =3 \ \  \ \ \mbox{ and} \ \
\delta_Z= \frac{3}{c^4_W} \left( \frac{7}{12} -\frac{10}{9}s_W^2+
\frac{40}{27}s_W^4 \right)
\end{eqnarray}
The factor  $\gamma_V=\Gamma_V^2/M_H^2$ accounts for the
non--zero width of the gauge boson $V$. The effect of the
finite width is important in the threshold region since
it guaranties the smooth transition from below-- to
above--threshold decays.
Again, $x_i$ are the scaled energies of the  fermions
in the final state.
The integration over the energies $x_1$ and $x_2$ between the boundaries in
eq.(\ref{boundary}) with $\kappa_{\Phi}$ replaced by $\kappa_V$, leads to
 the well--known result of Ref.\cite{R8}
for the partial width $[M_H \lsim 2M_V-\Gamma_V$]
\begin{eqnarray}
\Gamma (H \ra VV^*) &=& K_{HVV} ~\left[
\frac{1-8 \kappa_V+20\kappa_V^2}{(4\kappa_V-1)^{1/2}}
\arccos \left( \frac{3\kappa_V-1} {2\kappa_V^{3/2}} \right) \right. \non \\
&& \left. -\frac{1-\kappa_V}{6\kappa_V} (2-13\kappa_V+47\kappa^2_V) -
\frac{1}{2}(1-6\kappa_V+4\kappa_V^2) \log \kappa_V \right]
\end{eqnarray}
\s

\nn {\it (ii) Cascade decays}

\bigskip

Potentially interesting decay modes are cascade decays of the heavy
Higgs boson $H$ to mixed pairs of lighter Higgs bosons $h$, $A$ and
gauge bosons, $H\ra AZ$ and $H\ra H^{\pm}W^{\mp}$. Because
of the large value of the top quark mass the radiative corrections
will shift the $H$ mass upward for a given value of $M_A$ or
$M_{H^{\pm}}$ so that the phase space eventually opens for these decay modes.
The partial decay widths for the two--body decays above threshold
are given by the well-known expressions
\begin{eqnarray}
\Gamma(H \ra AZ) &=& \frac{G_F}{8\sqrt{2} \pi} \sin^2 (\beta-\alpha) \ \frac{
M_Z^4}{M_H} \lambda^\frac{1}{2}(M_A^2,M_Z^2;M_H^2) \lambda(M^2_A,M^2_H;
M^2_Z) \\
\Gamma(H \ra H^\pm W^\mp) &=& \frac{G_F}{8\sqrt{2} \pi} \sin^2 (\beta-\alpha)
\ \frac{M_W^4}{M_H} \lambda^\frac{1}{2}(M^2_{H^\pm},M^2_W;M^2_H)
\lambda(M^2_{H^\pm},M^2_H;M_W^2) \hspace*{4mm}
\end{eqnarray}
with $\lambda$ being the two--body phase space function
defined in eq.(\ref{lamb}).
Below--threshold decays of these modes are associated with virtual gauge
boson decays to light fermions [$\neq t$, see later], $H\ra AZ^*\ra Af\bar{f}$
and $H\ra H^{\pm}W^{\mp *} \ra H^{\pm}f\bar{f}'$, cf. Fig.~3. In analogy to
the gauge boson pair decays discussed in the previous paragraph, the
Dalitz plot densities for the two processes may be written
\begin{eqnarray}
\frac{ \dx \Gamma }{\dx  x_1 \dx x_2} (H \ra AZ^*\ra A f\bar{f}) &=&
 \frac{3 G_F^2 M_W^4}{8 \pi^3}
 \sin^2(\alpha-\beta)  M_H \delta_Z F_{AZ}(x_1, x_2)\\
\frac{ \dx \Gamma }{\dx  x_1 \dx x_2} (H \ra H^{\pm}W^{\mp *} \ra
H^\pm f\bar{f'}) &=&
 \frac{3 G_F^2 M_W^4}{8 \pi^3}
 \sin^2(\alpha-\beta) M_H \delta_W F_{H^{\pm}W}(x_1, x_2)
\end{eqnarray}
where the density function $F_{ij}$ is given by
\begin{eqnarray}
F_{ij}(x_1,x_2) = \frac{(1-x_1)(1-x_2)
-\kappa_i}{(1-x_1-x_2- \kappa_i +\kappa_j)^2 +\kappa_j\gamma_j} \label{funf}
\end{eqnarray}
Integrated over the Dalitz plots, the partial widths for decays sufficiently
below the thresholds  follow from
\beq
\Gamma (H \ra AZ^*\ra A f\bar{f}) &=&
 \frac{3 G_F^2 M_W^4}{8 \pi^3}
 \sin^2(\alpha-\beta)  M_H \delta_Z G_{AZ} \\
\Gamma (H \ra H^{\pm}W^{\mp *}\ra  H^\pm f\bar{f'}) &=&
 \frac{3 G_F^2 M_W^4}{8 \pi^3}
 \sin^2(\alpha-\beta)  M_H \delta_W G_{H^{\pm}W}
\end{eqnarray}
with
\begin{eqnarray}
G_{i j} =
\frac{1}{4} & & \hspace*{-5mm} \left\{ 2 (-1+\kappa_j - \kappa_i) \sqrt{
\lambda_{ij} } \ \left[ \frac{\pi}{2}+ { \arctan} \left( \frac{ \kappa_j (1-
\kappa_j +\kappa_i) -\lambda_{ij}} {(1- \kappa_i) \sqrt{\lambda_{ij}}} \right)
\right] \right. \nonumber \\
& & \left. + (\lambda_{ij} - 2\kappa_i) \log(\kappa_i) + \frac{1}{3}
(1-\kappa_i) \left[ 5(1+\kappa_i) -4 \kappa_j -\frac{2}{\kappa_j}
\lambda_{ij} \right] \right\} \label{fung}
\end{eqnarray}
and
\begin{eqnarray}
\lambda_{ij} = -1 + 2\kappa_i + 2 \kappa_j - (\kappa_i-\kappa_j)^2
\end{eqnarray}

\nn {\it (iii) Top decays}

\bigskip

Another class of important decays involves the top quark.
Above the top--quark threshold the partial decay width is given by
\begin{eqnarray}
\Gamma(H \ra \bar{t} t) & = &  \frac{3 G_F m_t^2}{4\sqrt{2} \pi}
\ g_{Htt}^2 \ M_{H} \ \beta^{3}_t
\end{eqnarray}
with $\beta_t=(1- 4m_t^2/M_H^2)^{1/2}$ being the velocity of the quarks in the
final state; the coupling $g_{Htt}=\sin\alpha/
\cos\beta$ has been discussed in the previous section [cf. Table 1
and Fig.~2a]. If the mass of the top quark in the Yukawa coupling is
evaluated at the scale of the Higgs mass $m_t(M_H)$, the bulk of the QCD
radiative corrections is effectively taken into account. Below the $tt$
threshold the final state $tbW$ can be reached through the $tt^*$
channel, Fig.~3. Since both the $W$ gauge boson mass and the top
quark mass must be kept non--zero\footnote{Since $m_b^2/M_H^2\ll 1$, the $b$
quark mass can be neglected in the amplitude but we shall keep it
non--zero in the phase space integration where its
effect can be important.},
the Dalitz plot density is more involved than in the previous cases.
Using  the more convenient variables $y_{1,2} =1-x_{1,2}$,  with $x_{1,2}
=2E_{t,b}/M_H$ for the reduced energies of the final top and bottom  quarks,
respectively, and the reduced masses $\kappa_i =M_i^2/M_H^2$,  one obtains
the Dalitz plot density
\begin{eqnarray}
\frac{ \dx \Gamma }{\dx x_1 \dx x_2} (H \ra W^- t\bar{b}) = \frac{3 G_F^2}
{64 \pi^3 } m_t^2 M_H^3 \frac{\sin^2 \alpha }{\sin^2\beta}\frac{\Gamma_{0} }
{ y_1^2+ \gamma_t \kappa_t}
\end{eqnarray}
The amplitude squared $\Gamma_{0}$ is given by
\begin{eqnarray}
\Gamma_{0}& =& y_1^2(1-y_1-y_2+\kappa_W-5\kappa_t) +2 \kappa_W (y_1 y_2-
\kappa_W -2\kappa_t y_1 +4 \kappa_t \kappa_W )  \non \\
&& - \kappa_t y_1 y_2 + \kappa_t(1-4 \kappa_t)(2 y_1 + \kappa_W+\kappa_t)
\eeq
When performing the numerical integration over the energies $x_1, x_2$
of the Dalitz plot, bounded by
\beq
\left| \frac{2(1-x_1-x_2+\kappa_t+\kappa_b-\kappa_W)+x_1 x_2}
            {\sqrt{x_1^2-4\kappa_t}\sqrt{x_2^2-4\kappa_b}} \right| \leq 1
\eeq
to obtain the partial decay widths, the non--zero decay width
of the virtual particles $\gamma_t=\Gamma^2_t/M_H^2$, as well as a non--zero
value for $\kappa_b=m_b^2/M_H^2$, have been included.
The charge conjugate $W^+
\bar{t} b$ final state doubles the partial width. \s

Finally,  partial widths for $H$ decays into two light
Higgs bosons have to be included, see eq.(\ref{HAA}).  For
small values of $\tb$, one needs to consider only the case where $\Phi=h,A$
are on--shell, since the $\Phi bb$ coupling is very small. \s

The branching ratios\footnote{The Fortran code for all partial decays widths of
the ${\cal SM}$ and ${\cal MSSM}$ Higgs bosons may be obtained from
djouadi@desy.de, jan.kalinowski@fuw.edu.pl or spira@desy.de.}  for the main
decay channels of the heavy Higgs boson $H$ are displayed in Fig.~6b for
$\tb=1.5$. Below the $t\bar{t} $ threshold, the decay $H\ra hh$ is the dominant
channel, superseded by decays to top quarks above the threshold. This rule is
only broken for Higgs masses of about 140 GeV where an accidentally small value
of the $\lambda_{Hhh}$ coupling allows the $b \bar{b}$ [no mix] and $WW^*$
[max mix] decay modes to become dominant. Important decay
modes in general, below the $t\bar{t}$ threshold, are decays to pairs of gauge
bosons, and $b\bar{b}$ decays. In a restricted range of the $H$ mass, also
below--threshold $AZ^*$ and $H^{\pm}W^{\mp *}$ play a non--negligible role.

\subsubsection*{4.2 The pseudoscalar Higgs boson $A$}

Since the pseudoscalar Higgs boson $A$ does not couple to pairs of $W$ and $Z$
bosons, the main two--body decay channels are $b\bar{b}$ and $\tau\tau$ decays
and, in the mass range between about 150 GeV and 350 GeV, cascade decays to
$hZ$ final states\footnote{Since the pseudoscalar Higgs boson is always lighter
than $H$ and $H^{\pm}$ in the ${\cal MSSM}$, the two--body decays $A\ra HZ$
and $A\ra H^{\pm}W^{\mp}$ are kinematically forbidden.}. For small $\tb$,
two--gluon decays are competitive in a limited Higgs mass range around 350 GeV.

\nn {\it (i) Cascade decays}

\bigskip

Above the threshold, the partial width of the cascade decay  $A\ra hZ$ of the
pseudoscalar Higgs boson reads \begin{eqnarray} \Gamma(A \ra hZ) =
\frac{G_F}{8\sqrt{2} \pi} \cos^2 (\beta-\alpha) \ \frac{M_Z^4}{M_A}
\lambda^{1/2}(M^2_h,M^2_Z; M_A^2) \lambda(M_h^2, M_A^2; M_Z^2) \end{eqnarray}
with $\lambda(x,y;z)$ being the two--body phase space function defined in
eq.(\ref{lamb}).  The branching ratio is sizeable for masses between the $hZ$
and the $t\bar{t}$ thresholds for  small values of $\tb$. \s

For $A$ masses below the threshold, three-body decays $A\ra hZ^* \ra
hf\bar{f}$, mediated by virtual $Z$ bosons, can play a role, Fig.~3. In the
same notation as before, the Dalitz plot density is given by
\begin{eqnarray}
\frac{ \dx \Gamma }{\dx  x_1 \dx x_2} (A \ra hZ^* \ra  h f\bar{f}) =
\frac{3 G_F^2 M_W^4}{8 \pi^3} \cos ^2(\alpha-\beta)
 M_A \delta_Z \ F_{hZ}(x_1,x_2)
\end{eqnarray}
with all possible $Z$ boson decays to light fermions summed up;
the Dalitz plot function $F$ is given in eq.(\ref{funf}).
Integrating out the energies $x_1$ and $x_2$ of the fermions, the partial
width of this decay process reads [$M_A <M_h+M_Z -\Gamma_Z$]
\begin{eqnarray}
\Gamma (A \ra hZ^* \ra  h f\bar{f}) =
\frac{3 G_F^2 M_W^4}{8 \pi^3} \cos ^2(\alpha-\beta)
M_A \delta_Z G_{hZ}
\end{eqnarray}
with $G$ defined  in eq.(\ref{fung}).

\bigskip

\nn {\it (ii) Top decays}

\bigskip

The three--body process $A\ra t\bar{b}W^-$, evolving out of the decay
\beq
\Gamma(A\ra t\bar{t})=\frac{3 G_F m_t^2}{4\sqrt{2} \pi} g^2_{Att} M_A \beta_t
\eeq
is predominantly mediated by virtual $t$ quarks,
$A\ra t\bar{t}^*$ followed by $\bar{t}^*
\ra \bar{b}W^-$. Using the same notation as for the process $H\ra
 t\bar{b}W^-$ we obtain for the Dalitz plot density
\begin{eqnarray}
\frac{ \dx \Gamma }{\dx x_1 \dx x_2} (A \ra  t\bar{b}W^-) =
\frac{3 G_F^2 m_t^2 M_A^3}{64 \pi^3 \tg^2 \beta}
 \frac{\Gamma_{0}}{y_1^2 +\kappa_t \gamma_t}
\end{eqnarray}
with
\begin{eqnarray}
\Gamma_{0} = y_1^2(1-y_1-y_2+\kappa_W-\kappa_t) +2 \kappa_W (y_1 y_2-
\kappa_W) - \kappa_t (y_1 y_2 -2y_1-\kappa_W-\kappa_t)
\end{eqnarray}
This reduced density approaches the corresponding expression for the
scalar CP-even $H$ decay $H\ra t\bar{b}W^-$ in the limit $M_{H,A} \gg m_t$ as
expected from chiral symmetry. The non--zero width $\gamma_t=m_t^2/M_A^2$ of
the top quark and the finite $b$--quark mass are taken into account when the
Dalitz plot density is integrated out to obtain the partial width. \s

The numerical results for the two-- and three--body decays of the
pseudoscalar Higgs boson $A$ are summarized in Fig.~6c. The dominant
modes are the $H\ra b\bar{b}$ and $H \ra t\bar{t}$ decays below the
$hZ$ and $t\bar{t}$ thresholds respectively; in the intermediate
mass region, $M_A=200$ to $300$ GeV, the decay $H \ra AZ^*$ [which
reaches the percent level already at $M_A =130$ GeV] dominates. Note that
this decay mode is more important in the case of no mixing. The gluonic
decays are significant around the $t\bar{t}$ threshold.

\subsubsection*{4.3 The charged Higgs boson $H^{\pm}$}

Except for the intermediate mass region, the main decay channels of the charged
Higgs particles are the fermionic $\tau\nu_{\tau}$ and $tb$ modes. The bulk of
the radiative QCD corrections can be taken into account by using running quark
masses. Besides these channels, $H^{+} \ra c\bar{s}$ and $c\bar{b}$ modes are
non--negligible, with the CKM--type suppression of the $cb$ current
[$V_{cb} \sim 0.04$] partly
compensated by the overall coupling $\sim m_b$ of this current.

\newpage

\nn {\it (i) Cascade decays}

\bigskip

If allowed kinematically, charged Higgs bosons can decay into the lightest
neutral Higgs boson plus a $W$ boson
\begin{eqnarray}
\Gamma(H^{+} \rightarrow hW^-) =
\frac{G_F \cos^2 (\beta-\alpha) }{8\sqrt{2}\pi
c_W^2} \frac{M_W^4}{M_{H^\pm}} \lambda^{\frac{1}{2}}(M_h^2,M_W^2;M_{H^{\pm}}^2)
\lambda(M_h^2,M_{H^{\pm}}^2;M_W^2) \hspace*{0.2cm}
\end{eqnarray}
The branching ratio is quite substantial below the top threshold for small
values of $\tb$ for which the $H^{\pm}W^{\mp}h$ coupling is not suppressed. The
two--body decay of $H^+$ to $W^+A$ is kinematically not allowed. \s

There are three below--threshold decays of the charged Higgs boson
with substantial branching ratios, two of which involve a virtual $W$
boson, $H^{\pm} \ra hW^{\pm *}\ra h f\bar{f}'$ and $H^{\pm}\ra AW^{\pm *}
\ra Af \bar{f}'$, and one involves a virtual $t$ quark,
$H^+\ra \bar{b}t^*\ra \bar{b}bW^+$, Fig.~3. \s

The Dalitz plot densities involving the virtual $W$ boson are given by
\begin{eqnarray}
\frac{ \dx \Gamma }{\dx  x_1 \dx x_2} (H^+ \ra
h W^{+ *}\ra h f\bar{f})& =& \frac{9 G_F^2 M_W^4}
{8 \pi^3} \cos ^2(\alpha-\beta) M_{H^\pm} \ F_{hW}(x_1, x_2) \\
\frac{ \dx \Gamma }{\dx  x_1 \dx x_2} (H^+\ra  A W^{+ *}
  \ra A f\bar{f})& =& \frac{9 G_F^2 M_W^4}
{8 \pi^3} M_{H^\pm} \ F_{AW}(x_1, x_2)
\end{eqnarray}
[$\kappa_i = M_i^2/M_{H^\pm}^2$ and $x_{1,2}$ being the energies of the $f
\bar{f}$ final states], from which  one obtains the partial decay widths
below the thresholds
\begin{eqnarray}
\Gamma (H^+\ra h W^{+ *} \ra h f\bar{f})& =& \frac{9 G_F^2 M_W^4}{8 \pi^3}
\cos ^2(\alpha-\beta) M_{H^\pm} \ G_{h W} \\
\Gamma (H^+ \ra A W^{+ *}\ra  A f\bar{f})& = & \frac{9 G_F^2 M_W^4} {8 \pi^3}
M_{H^\pm} \ G_{AW}
\end{eqnarray}
The coefficients $F$ and $G$ have been defined in eqs.(\ref{funf},
\ref{fung}).

\bigskip

\nn {\it (ii) Top decays}

\bigskip

For the Dalitz plot density of the decay mode $H^+\ra \bar{b}t^* \ra
\bar{b}bW^+$, evolving out of
\beq
\Gamma(H^+\ra t\bar{b})&=&\frac{3 G_F}{4\sqrt{2} \pi} \frac{\lambda^{1/2}
(t,b,H^{\pm})}{M_{H^{\pm}}} |V_{tb}|^2
\times \non \\
& & \left[ (M_{H^\pm}^{2} -m_{t}^{2}-m_{b}^{2}) \left( m_{b}^{2} {\tg}^2
\beta + m_t^2{\ctg}^2 \beta \right) -4m_t^2m_b^2 \right]
\end{eqnarray}
[with  the CKM--type  matrix element $V_{tb} \simeq 1$], we obtain
\begin{eqnarray}
\frac{ \dx \Gamma }{\dx  x_1 \dx x_2} (H^+ \ra
\bar{b}t^*\ra W b\bar{b}) = K_{H^\pm tb} \
\frac{(1-x_1)(1-x_2)/\kappa_W + 2x_1 +2 x_2 -3 +2\kappa_W}{(1-x_2- \kappa_t)^2
+\kappa_t \gamma_t }
\end{eqnarray}
with
\begin{eqnarray}
K_{H^{\pm} tb} = \frac{3 G_F^2 m_t^4}{32 \pi^3 }
\frac{1}{ {\tg}^2 \beta} \ M_{H^\pm}
\end{eqnarray}
in the limit where the $b$ quark mass can be neglected; $x_{1,2}=2
E_{b,\bar{b}}/M_{H^{\pm}}$. Integrating out the energies of the
Dalitz plot, the partial width [$M_{H^\pm} <m_t+m_b - \Gamma_t$]
can be written as
\beq
\Gamma (H^\pm \ra W b\bar{b}) = \frac{1}{2} K_{H^\pm tb} \hspace*{-4mm} & &
\left\{\frac{ \kappa_W^2}{\kappa_t^3} (4 \kappa_W \kappa_t+3\kappa_t-4\kappa_W)
\log \frac{ \kappa_W (\kappa_t -1) }{\kappa_t - \kappa_W} \right. \\
& & +(3\kappa_t^2 -4\kappa_t -3\kappa_W^2+1) \log \frac{ \kappa_t -1}{\kappa_t
- \kappa_W} -\frac{5}{2} \non \\
& & \left. + \frac{1- \kappa_W}{\kappa_t^2} ( 3\kappa_t^3 -\kappa_t \kappa_W
-2\kappa_t \kappa_W^2+4\kappa_W^2) + \kappa_W \left( 4 -\frac{3}{2} \kappa_W
\right) \right\} \non
\end{eqnarray}

The branching ratios of the most important two-- and three--body decays,
including the threshold effects in the numerical analysis, are displayed in
Fig.~6d. The branching ratio is reduced for the $\tau\nu$ search channel of the
charged Higgs boson quite significantly; indeed, by including of the
three--body decays, this decay does not overwhelm all the other modes since the
$H^+ \ra hW^*$ channel as well the channels $H\ra AW^*$ in the low mass range
and $H^+ \ra bt^*$ in the intermediate mass range, have appreciable branching
ratios. [Here, as well as for the decay $A \ra hZ^*$, we differ from
Ref.\cite{R9}.] It is interesting to observe that the mixing leads to a
qualitative change of the decay pattern in the intermediate mass range: in
the ``no mixing" scenario the $hW^*$ decay mode is much more important than
with mixing for moderately large charged Higgs masses.

\subsection*{5. Total widths of the Higgs particles}

It is well known that the total widths of the \SUSY Higgs particles are in
general considerably smaller than the width of the ${\cal SM}$ Higgs particle.
\s

The large width of the \SM\ Higgs boson for high masses is due to the decays to
longitudinal $W/Z$ bosons which grow as $G_F M_H^3$. The absence of the gauge
boson couplings to $A$ and the decoupling from $H$ for large  masses shut this
channel. The dominant decay modes are built-up by top quarks so that the widths
rise only linearly with the Higgs masses $\sim G_F m_t^2 M_H$. Detailed numbers
are shown for the two different mixing scenarios in Fig.~7. While the width of
the ${\SM}$ Higgs boson is of order $\Gamma_{H_{{\cal SM}}} \sim 200 $ GeV for
a Higgs mass $M_{H_{ {\cal SM}}}\sim 750 $ GeV, the width of the \SUSY Higgs
particles remain less than $\sim 10$ GeV in this mass range for moderate
values of $\tb$. \s

For large $\tb$ values, the decay widths of all the five Higgs bosons are
determined by $b$--quark finale states and they scale
like ${\rm tg}^2\beta$ [except for $h$ and $H$ near their maximal and minimal
mass values, respectively]. The $H,A$ and $H^+$ widths therefore become
experimentally significant, for $\tb$ values of order $\sim 50$ and above
and for large Higgs masses \cite{Janot}.

\newpage

\subsubsection*{Acknowledgements}
Discussions with M. Spira are gratefully acknowledged. We thank S. Moretti and
W.J. Stirling for a correspondence on pseudoscalar and charged Higgs decays.
We also thank M. Carena for providing us with a Fortran code for Higgs masses
and couplings in which the full dependence on the SUSY parameters $\mu$
and $A_t$ is taken into account.  A.D. thanks the Theory Group for the warm
hospitality extended to him at DESY.

\vspace*{3cm}

\subsection*{Appendix}
\renewcommand{\theequation}{A.\arabic{equation}}
\setcounter{equation}{0}

Two of the three-body decays discussed before, $H \ra tbW$ and $A \ra tbW$, can
be mediated by a superposition of different intermediate states. Even though
these effects are rather small, they should nevertheless be given in this
appendix for the sake of completeness. Omitting the finite decay widths of the
virtual particles (which can easily be implemented) and using the notation
introduced before, the Dalitz plot density may be written in the notation
introduced before,

\bigskip

\nn a) \underline{$H \ra \bar{t}t^* + W^+ H^{-*} +W^+ W^{-*}
\ra W^+ \bar{t}b$ :}

\begin{eqnarray}
&& \hspace*{-5mm}
\frac{ \dx \Gamma }{\dx x_1 \dx x_2} (H \ra W^- t\bar{b}) = \frac{3 G_F^2}
{64 \pi^3} m_t^2 M_H^3 \ \times  \\
&& \hspace*{-3mm}
\left[ \frac{\sin^2 \alpha ~\Gamma_{11} }{\sin^2 \beta ~y_1^2}+\frac{2
\sin\alpha \sin(\beta- \alpha) ~\Gamma_{12} }{\sin \beta \tb ~y_1(1-
y_1-y_2+ \kappa_W-\kappa_+)} + \frac{ \sin^2(\beta-\alpha) ~\Gamma_{22}}{
{\rm tg}^2 \beta ~(1-y_1-y_2 +\kappa_W-\kappa_+) ^2} \right. \non \\
&& \hspace*{-3mm}
\left. + \frac{ 2 \sin \alpha \cos(\beta-\alpha)~\Gamma_{13}}
{ \sin \beta ~y_1 (1-y_1-y_2)}  + \frac{ 2 \sin(\beta-\alpha) \cos
(\beta-\alpha) ~\Gamma_{23} }{ {\rm tg} \beta ~(1-y_1-y_2+\kappa_W-\kappa_+)
(1-y_1 -y_2)} + \frac{\cos^2 (\beta-\alpha) ~\Gamma_{33}}{ (1-y_1-y_2)^2}
\right] \non
\end{eqnarray}
with the reduced amplitudes squared $\Gamma_{ii}$ and the interference terms
$\Gamma_{ij}$ given by
\begin{eqnarray}
\Gamma_{11} &=& y_1^2(1-y_1-y_2+\kappa_W-5\kappa_t) +2 \kappa_W (y_1 y_2-
\kappa_W)  \non \\
&& +4 \kappa_t (2 \kappa_W^2 -\kappa_W \kappa_t -\kappa_t^2 -\kappa_W y_1
-2y_1 \kappa_t )+\kappa_t( 2y_1 -y_1 y_2 +\kappa_W+ \kappa_t) \non \\
\Gamma_{12} &=& (1-y_1-y_2 +\kappa_W)( 2\kappa_W -y_1^2- y_1y_2) + 2 \kappa_t
\kappa_W ( y_2 - y_1 -3) \non \\
&& + \kappa_t (y_1+y_2)(3y_1 +y_2-1 +2\kappa_t) \non \\
\Gamma_{22} &=& (y_1 +y_2-1- \kappa_W+\kappa_t) \left[ 4\kappa_W -
(y_1+y_2)^2 \right] \non \\
\Gamma_{13} &=& y_1^2 (1- \kappa_W -y_1- y_2-2\kappa_t) +y_1 y_2(1-y_1 -y_2
+3\kappa_W - 4\kappa_t) \non \\
&& +2\kappa_W^2(4\kappa_W-2y_2-3y_1) +2\kappa_W (y_2+y_1-1) \non \\
&& +\kappa_t \left[ y_2(1-y_2-2\kappa_t)+
y_1(1-y_1 -2\kappa_t)-2\kappa_W (2\kappa_W+2\kappa_t +2y_1-3) \right] \non \\
\Gamma_{23} &=& y_1^2 (\kappa_W -1 +y_1+3y_2+\kappa_t) +y_2^2(-\kappa_W
-1+y_2+3y_1+ \kappa_t) +2y_1 y_2\kappa_t \non \\
&& -2 y_1 y_2 +2\kappa_W\left[ y_2(\kappa_W-2+\kappa_t)+
y_1(-\kappa_W-2+\kappa_t)+2 (1-\kappa_t) \right] \non \\
\Gamma_{33} &=& 4 \kappa_W^2 \left[ y_1y_2+\kappa_W +2\kappa_W (\kappa_W
-y_1 -y_2)\right]/ \kappa_t \non \\
&& + y_1^2 ( 1- \kappa_t -3\kappa_W-y_1 -3y_2)
+  y_2^2 ( 1- \kappa_t + \kappa_W-y_2 -3y_1)  \non \\
&& + 2y_1y_2 (1-\kappa_t-\kappa_W) -4\kappa_W \left[ \kappa_W^2+
(\kappa_t -1) (y_1+ y_2 -1 + \kappa_W) \right]
\end{eqnarray}

\bigskip

\nn b) \underline{$A \ra \bar{t}t^* + W^+ H^{-*} W^{-*} \ra W^+ \bar{t}b$ :}

\begin{eqnarray}
\frac{ \dx \Gamma }{\dx x_1 \dx x_2} (A \ra W^- t\bar{b}) &&=
\frac{3 G_F^2}{64 \pi^3} \frac {m_t^2 M_A^3}
 {\tg ^2 \beta}  \times \\
&& \left[ \frac{\Gamma_{11}}{y_1^2}  + \frac{ 2
\Gamma_{12}} {y_1(1- y_1-y_2+\kappa_W-\kappa_+)} + \frac{\Gamma_{22}}{
(1-y_1-y_2+\kappa_W-\kappa_+) ^2} \right] \non
\end{eqnarray}
with the reduced amplitudes squared  and the interference term
\begin{eqnarray}
\Gamma_{11} &=& y_1^2(1-y_1-y_2+\kappa_W-\kappa_t) +2 \kappa_W (y_1 y_2-
\kappa_W) - \kappa_t (y_1 y_2 -2y_1-\kappa_W-\kappa_t) \non \\
\Gamma_{12} &=& (1-y_1-y_2 +\kappa_W)( 2\kappa_W -y_1^2- y_1y_2)
+ \kappa_t [ (y_1+y_2)^2 -y_1 -y_2-2 \kappa_W] \non \\
\Gamma_{22} &=& (y_1 +y_2-1- \kappa_W+\kappa_t) \left[ 4\kappa_W -
(y_1+y_2)^2 \right]
\end{eqnarray}

\bigskip

\nn As expected from chiral symmetry, the expressions $\Gamma_{ij}$ approach
the corresponding expressions for the decay $H \ra t\bar{b} W$ in the limit
$M_{H,A}\gg m_t$. Note that  eqs.(A2) and (A4) are to be multiplied  by a
factor of 2 if  the charge conjugate processes are taken into account.

\vspace*{3cm}


\newpage

\nn {\Large \bf Figure Captions}

\begin{itemize}

\item[{\bf Fig.~1:~}]
The masses of the \MSSM\ neutral CP--even Higgs bosons $h,H$
and the charged Higgs
boson $H^\pm$ as a function of the pseudoscalar $A$ mass for two values
of $\tg \beta=1.5$ (solid lines) and $\tg \beta=30$ (dashed lines); the
long-dashed line presents $M_{H^\pm}$. a) ``no mixing": $A_t=0$, $\mu=100$
GeV and $M_S=1$ TeV, and b) ``maximal mixing": $A_t=\sqrt{6}M_S, \mu=\ll M_S$
and $M_S=1$ TeV. [$M_{H^\pm}$ does not depend on the mixing.]

\item[{\bf Fig.~2a:}]
The coupling parameters of the neutral CP--even Higgs bosons to
fermions and gauge bosons as functions of the Higgs masses for
the two values $\tb=1.5$ and 30; no mixing (solid lines) and
maximal mixing (dashed lines). The couplings are
 normalized to the \SM\ couplings as defined in Table 1.

\item[{\bf Fig.~2b:}]
The trilinear couplings of the \MSSM\ heavy neutral CP--even Higgs boson.
(i) $\lambda_{Hhh}$ and (ii) $\lambda_{HAA}$ as functions of the $H$ mass
for the two values $\tb=1.5$ and 30; no mixing (solid lines)
and maximal mixing (dashed lines).

\item[{\bf Fig.~3:~}]
Main mechanisms for below--threshold three--body \MSSM\ Higgs decays.

\item[{\bf Fig.~4:~}]
The branching ratios of the \MSSM\ Higgs bosons $h$, $H$, $A$ and $H^\pm$ as
functions of their masses for $\tb=30$; no mixing
(solid lines) and maximal mixing (dashed lines). For the pseudoscalar and
charged Higgs bosons the two mixing scenarios are indistinguishable in
this case.

\item[{\bf Fig.~5a:}]
The branching ratios of the light CP--even \MSSM\ Higgs boson $h$ as a
function of its maximum mass value for the two
mixing scenarios. The dashed
band attached to the charm branching ratio, indicates the QCD uncertainties.

\item[{\bf Fig.~5b:}]
The branching ratios of the heavy CP--even \MSSM\ Higgs boson $H$ near its
minimum mass value as a function of $M_A$ for $\tb=30$ and for the two mixing
scenarios.

\item[{\bf Fig.~6a:}]
The branching ratios of the light CP--even \MSSM\ Higgs boson $h$ for
$\tb=1.5$.

\item[{\bf Fig.~6b:}]
The branching ratios of the heavy CP--even \MSSM\ Higgs boson $H$ for
$\tb=1.5$.

\item[{\bf Fig.~6c:}]
The branching ratios of the CP--odd \MSSM\ Higgs boson $A$ for $\tb=1.5$;
(i) no mixing and (ii) maximal mixing.

\item[{\bf Fig.~6d:}]
The branching ratios of the \MSSM\ charged Higgs boson $H^\pm$ for $\tb=1.5$;
(i) no mixing and (ii) maximal mixing.

\item[{\bf Fig.~7:~}]
The total widths of the four \MSSM\  Higgs bosons $h,H,A$ and
$H^\pm$ as functions
of the pseudoscalar mass $M_A$ for the two values $\tb=1.5$ and 30/50 and  the
two scenarios of ``no mixing" and ``maximal mixing".

\end{itemize}

\newpage

\begin{figure}[htbp]
\vspace*{1cm}
\centerline{\psfig{figure=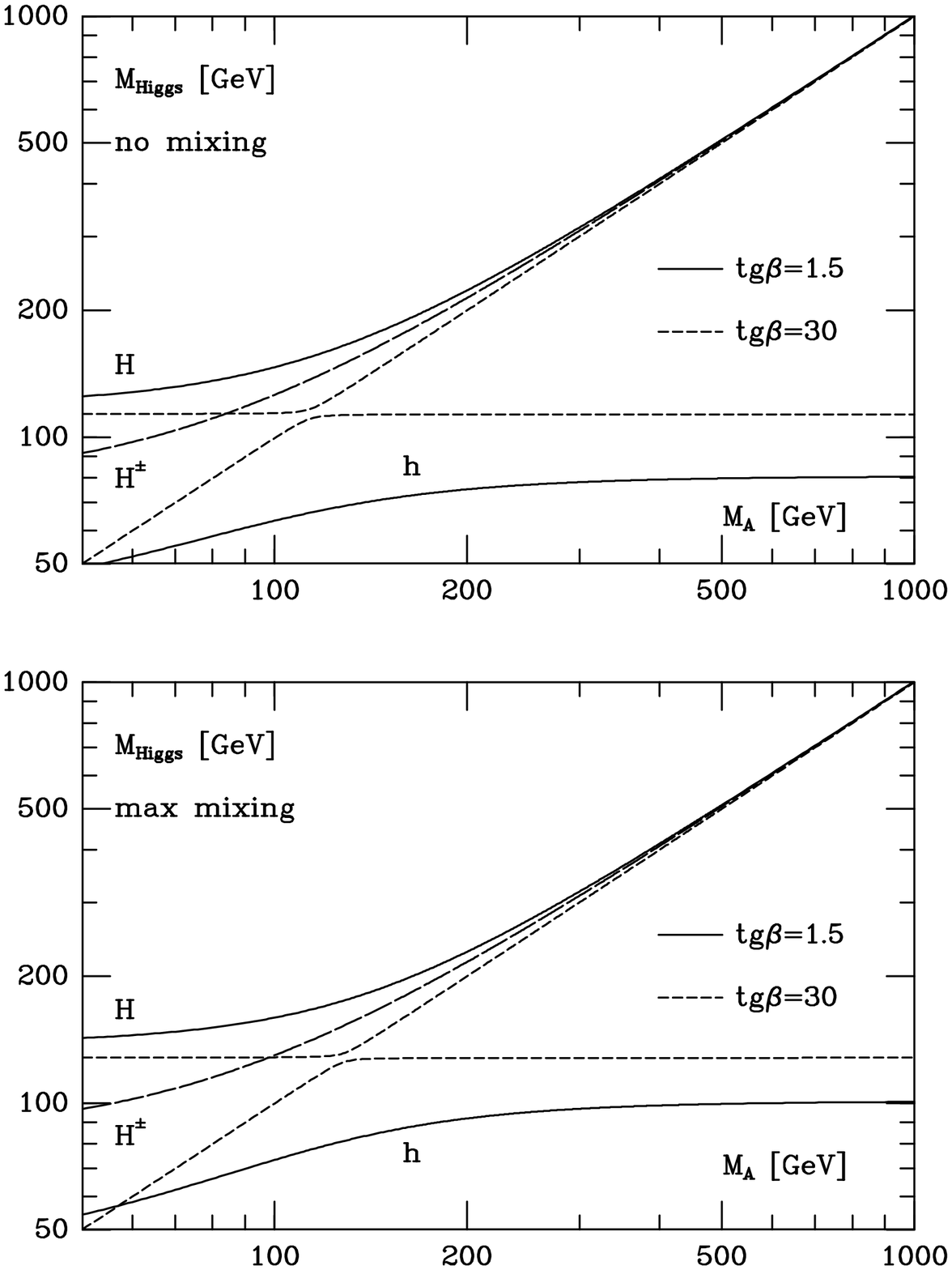,height=21.5cm,width=15cm}}
\vspace*{-3cm}
\centerline{\bf Fig.~1}
\end{figure}

\newpage

\begin{figure}[htbp]
\vspace*{1cm}
\centerline{\psfig{figure=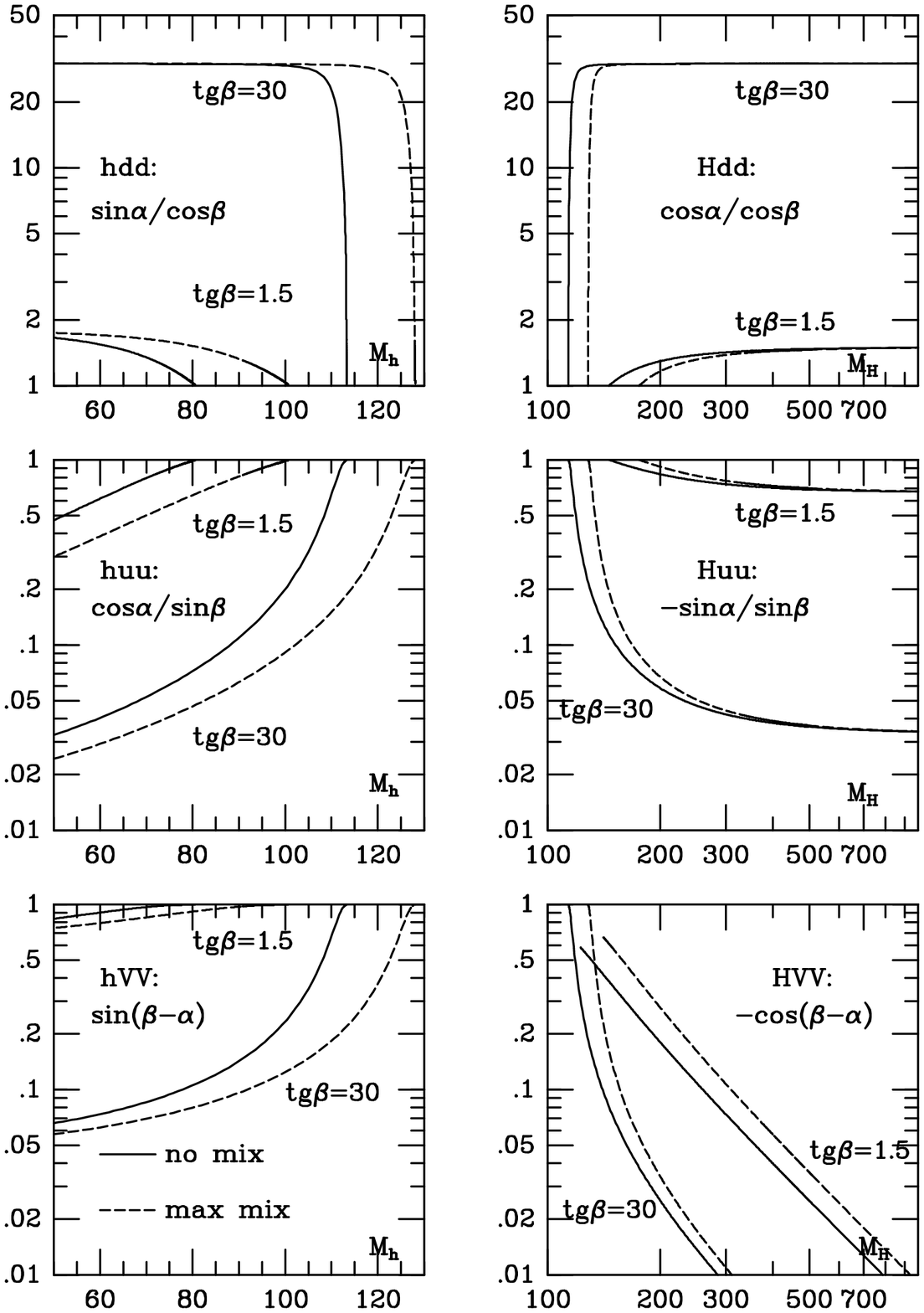,height=21.5cm,width=15cm}}
\vspace*{-3cm}
\centerline{\bf Fig.~2a}
\end{figure}

\newpage

\begin{figure}[htbp]
\vspace*{1cm}
\centerline{\psfig{figure=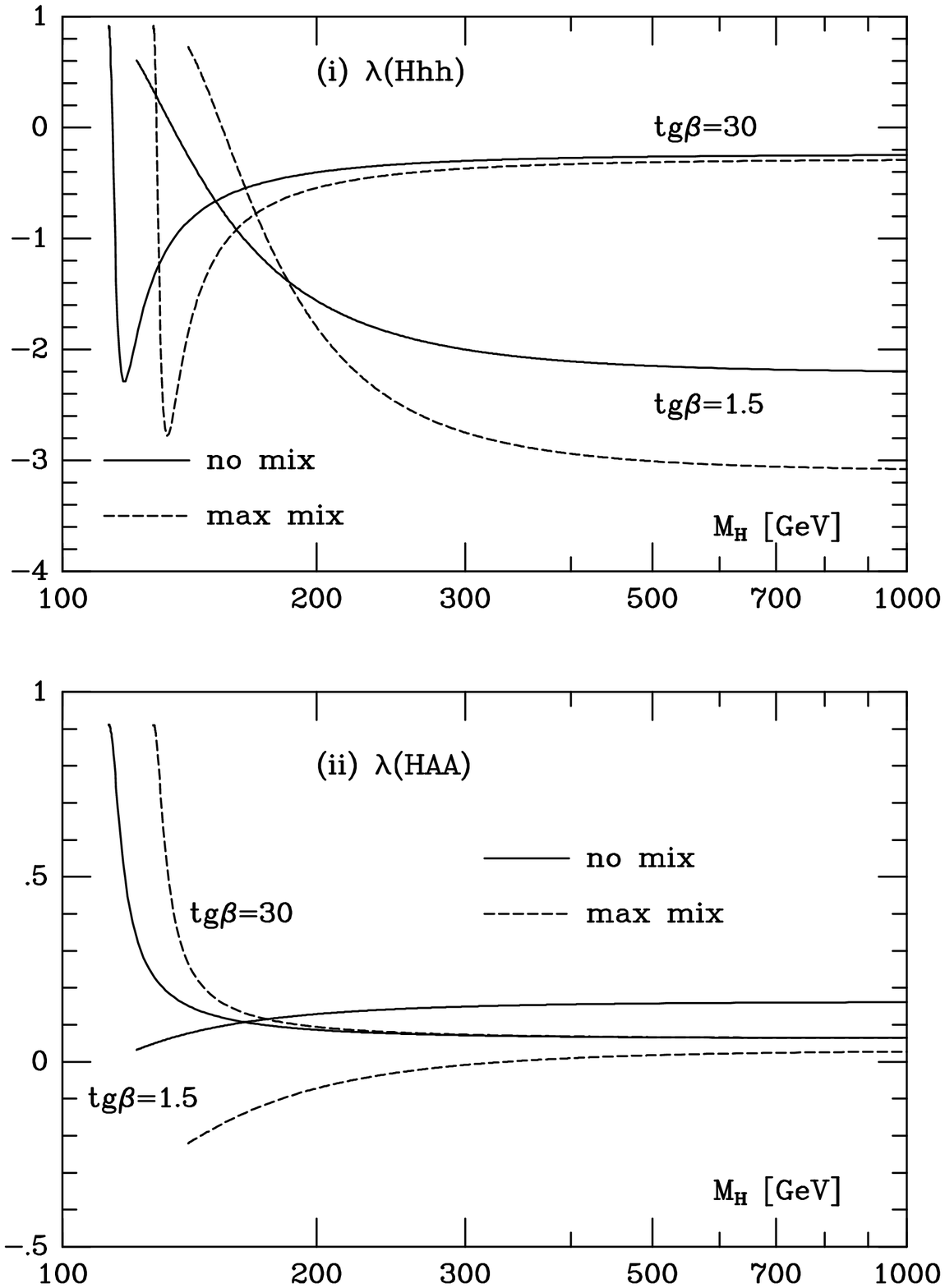,height=21.5cm,width=15cm}}
\vspace*{-3cm}
\centerline{\bf Fig.~2b}
\end{figure}

\newpage

\begin{figure}[htbp]
\vspace*{1cm}
\centerline{\psfig{figure=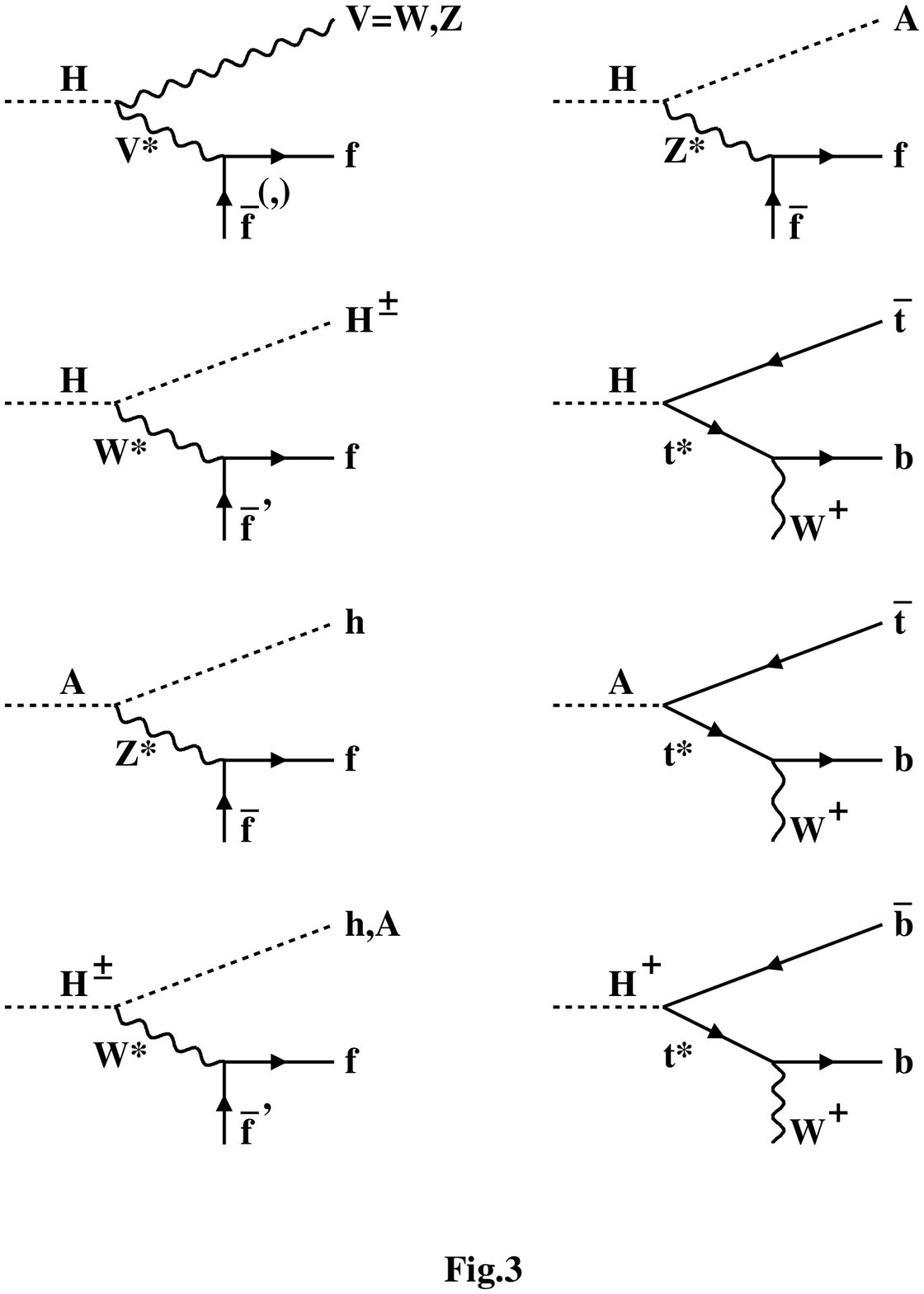,height=20.5cm,width=15cm}}
\end{figure}

\newpage

\begin{figure}[htbp]
\vspace*{1cm}
\centerline{\psfig{figure=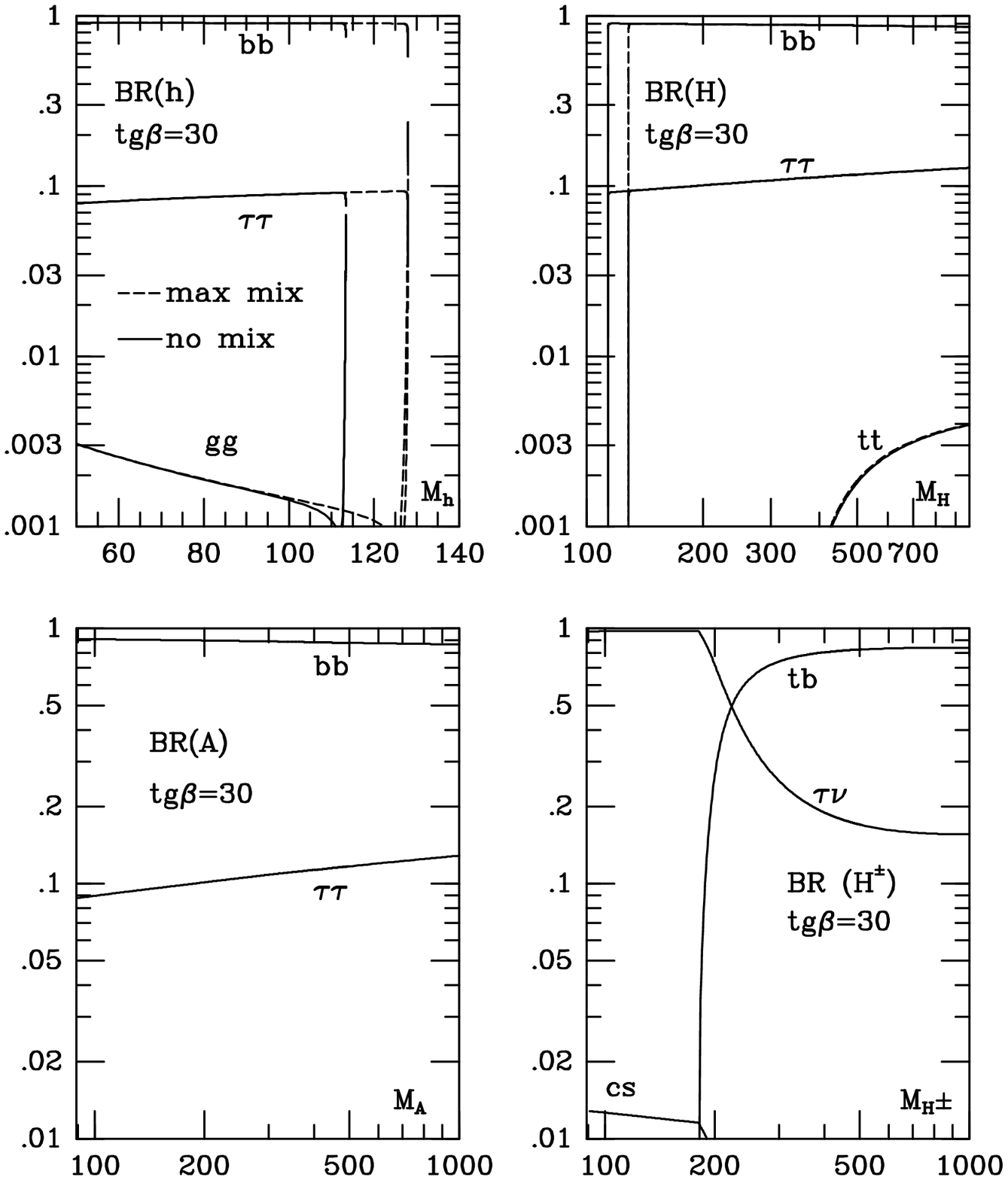,height=21.5cm,width=15cm}}
\vspace*{-3cm}
\centerline{\bf Fig.~4}
\end{figure}

\newpage

\begin{figure}[htbp]
\vspace*{1cm}
\centerline{\psfig{figure=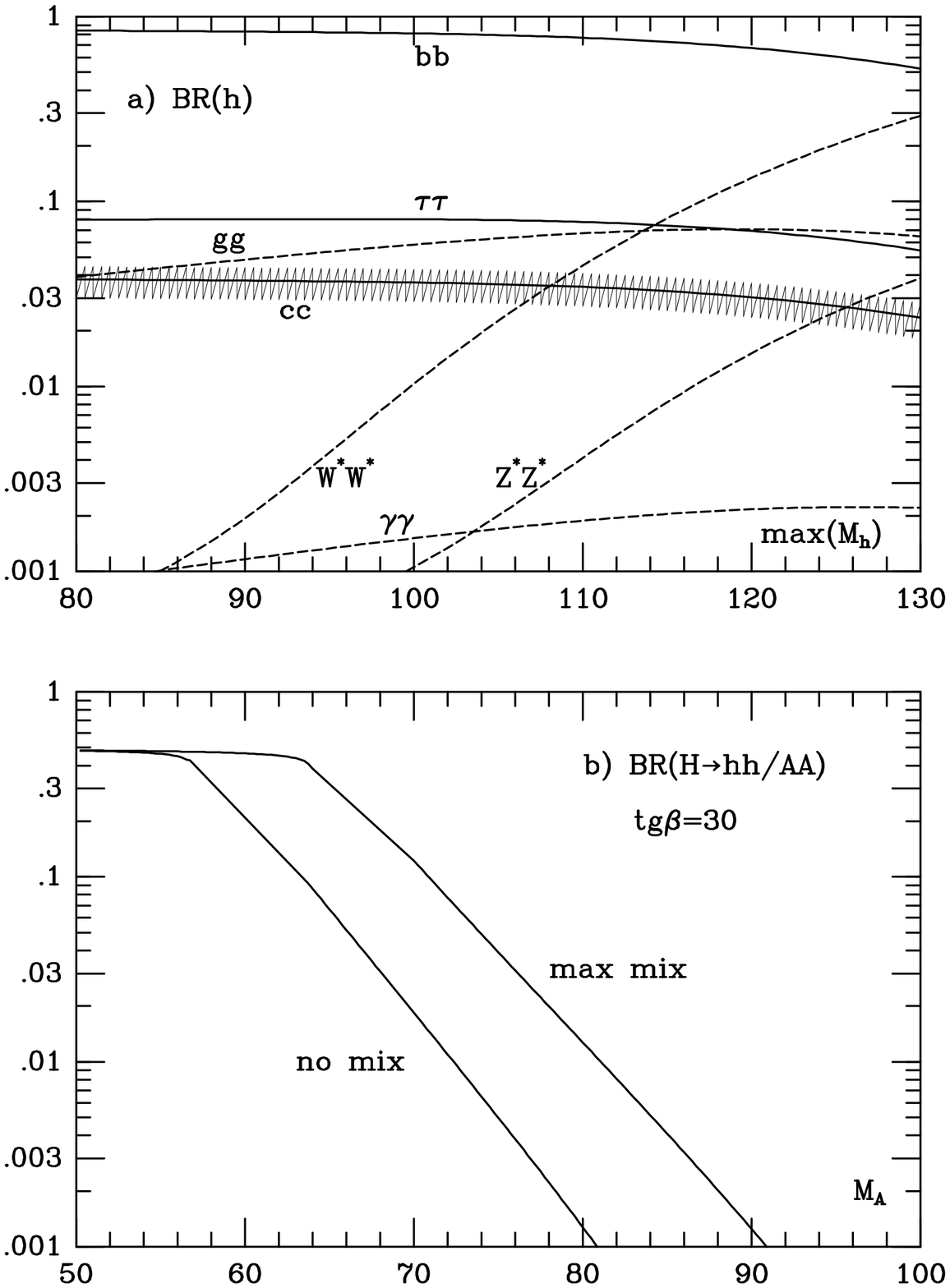,height=21.5cm,width=15cm}}
\vspace*{-3cm}
\centerline{\bf Fig.~5}
\end{figure}

\newpage

\begin{figure}[htbp]
\vspace*{1cm}
\centerline{\psfig{figure=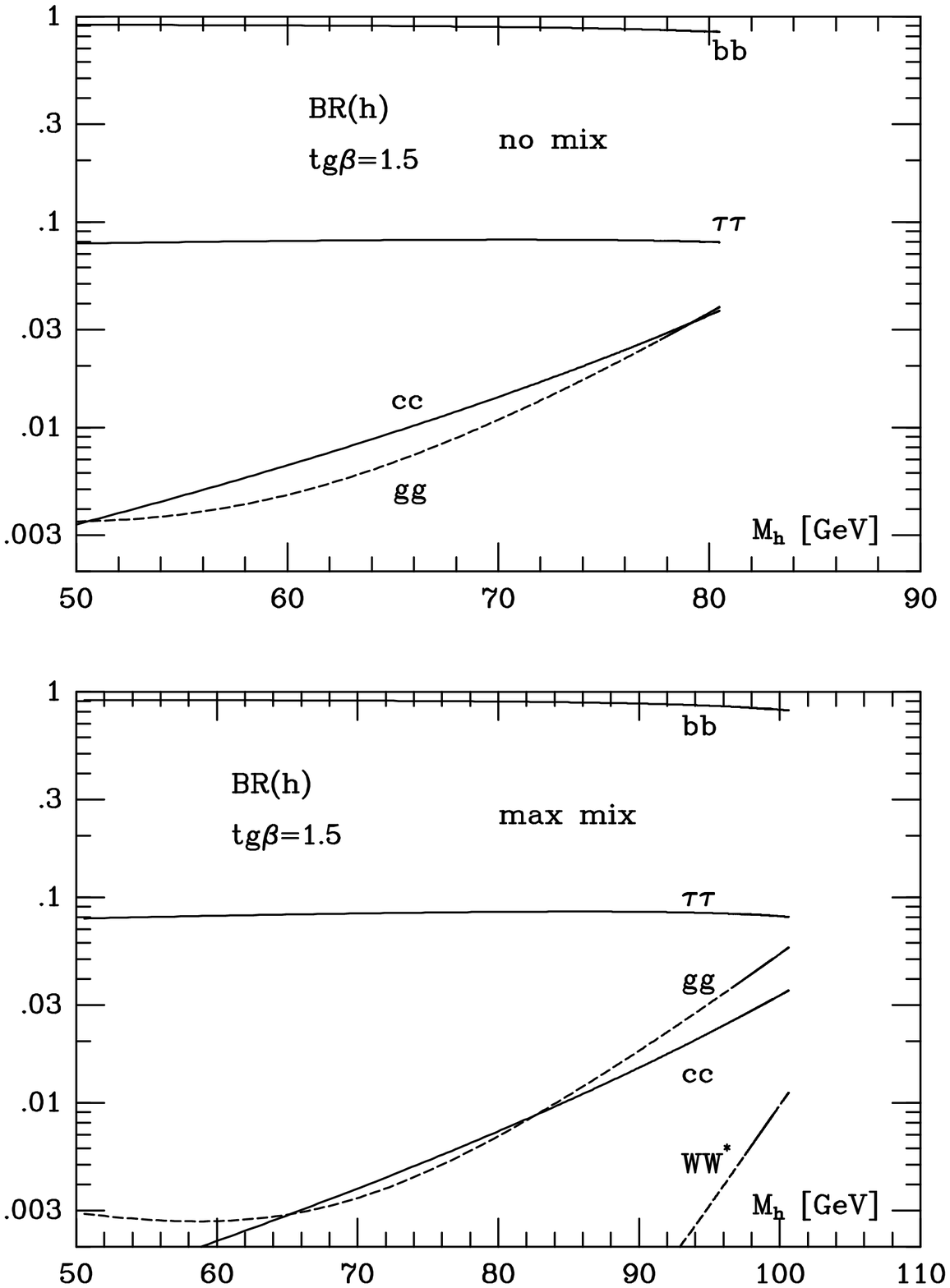,height=21.5cm,width=15cm}}
\vspace*{-3cm}
\centerline{\bf Fig.~6a}
\end{figure}

\newpage

\begin{figure}[htbp]
\vspace*{1cm}
\centerline{\psfig{figure=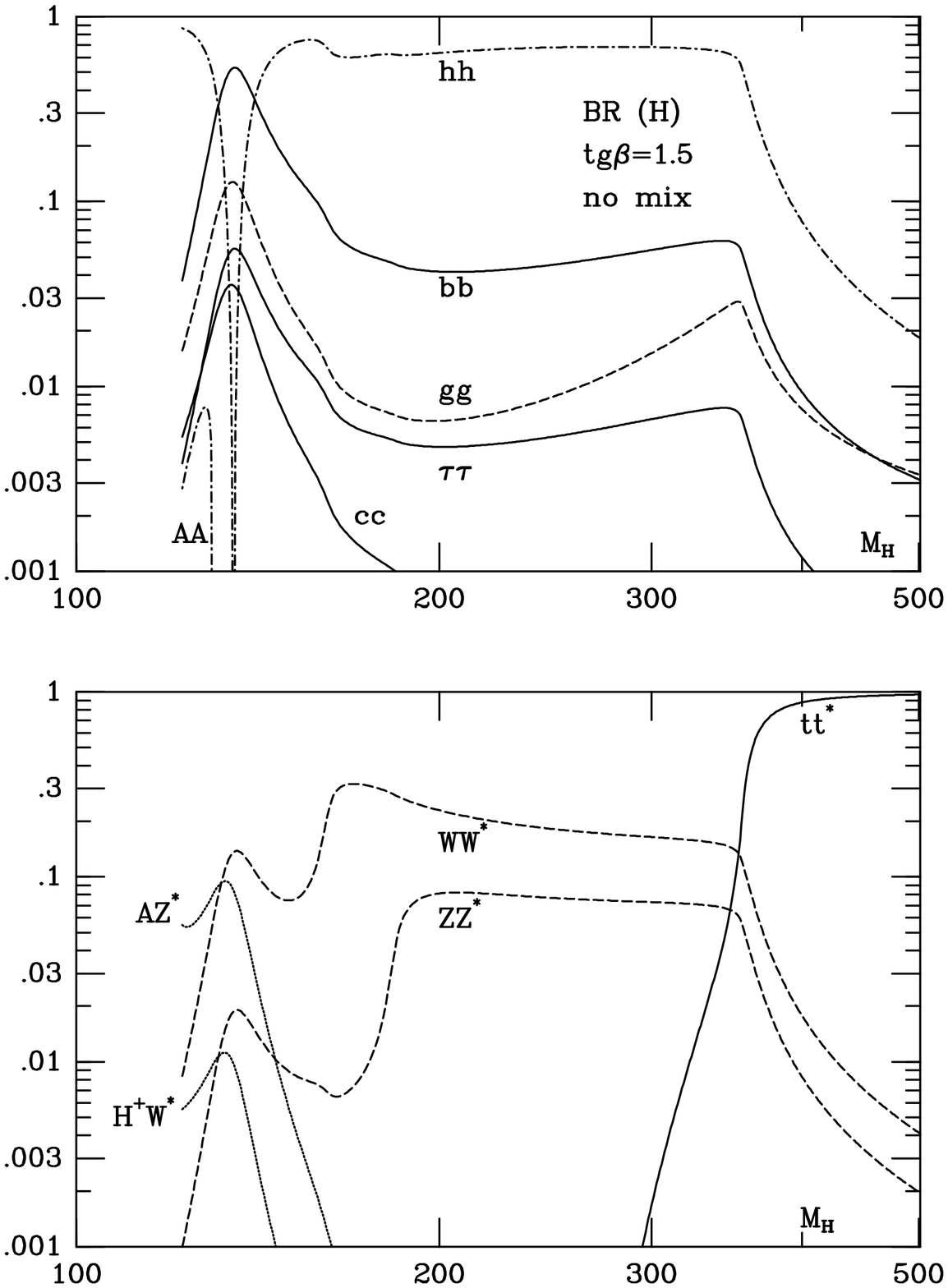,height=21.5cm,width=15cm}}
\vspace*{-3cm}
\centerline{\bf Fig.~6b}
\end{figure}

\newpage

\begin{figure}[htbp]
\vspace*{1cm}
\centerline{\psfig{figure=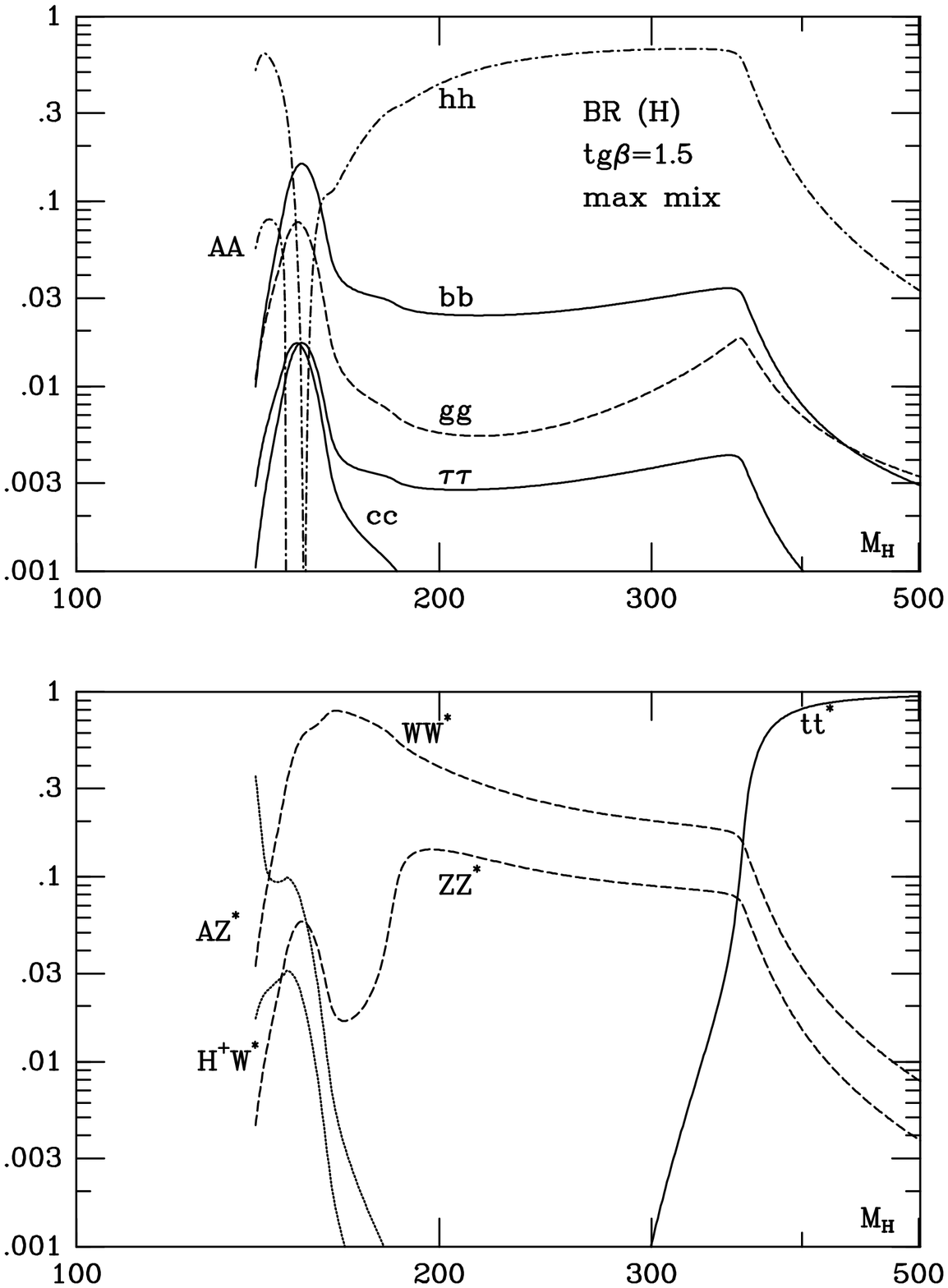,height=21.5cm,width=15cm}}
\vspace*{-3cm}
\centerline{\bf Fig.~6b (cont.)}
\end{figure}

\newpage

\begin{figure}[htbp]
\vspace*{1cm}
\centerline{\psfig{figure=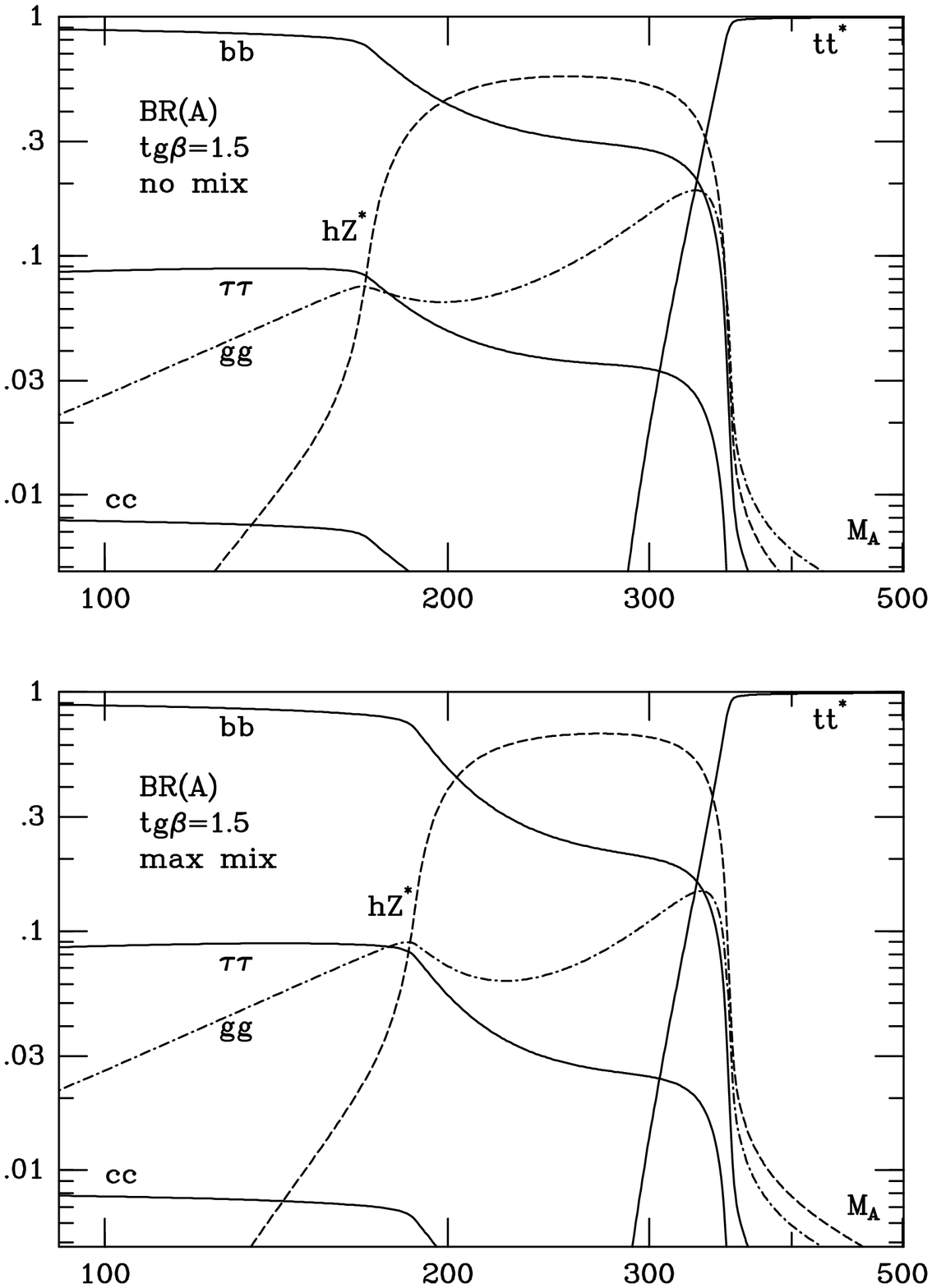,height=21.5cm,width=15cm}}
\vspace*{-3cm}
\centerline{\bf Fig.~6c}
\end{figure}

\newpage

\begin{figure}[htbp]
\vspace*{1cm}
\centerline{\psfig{figure=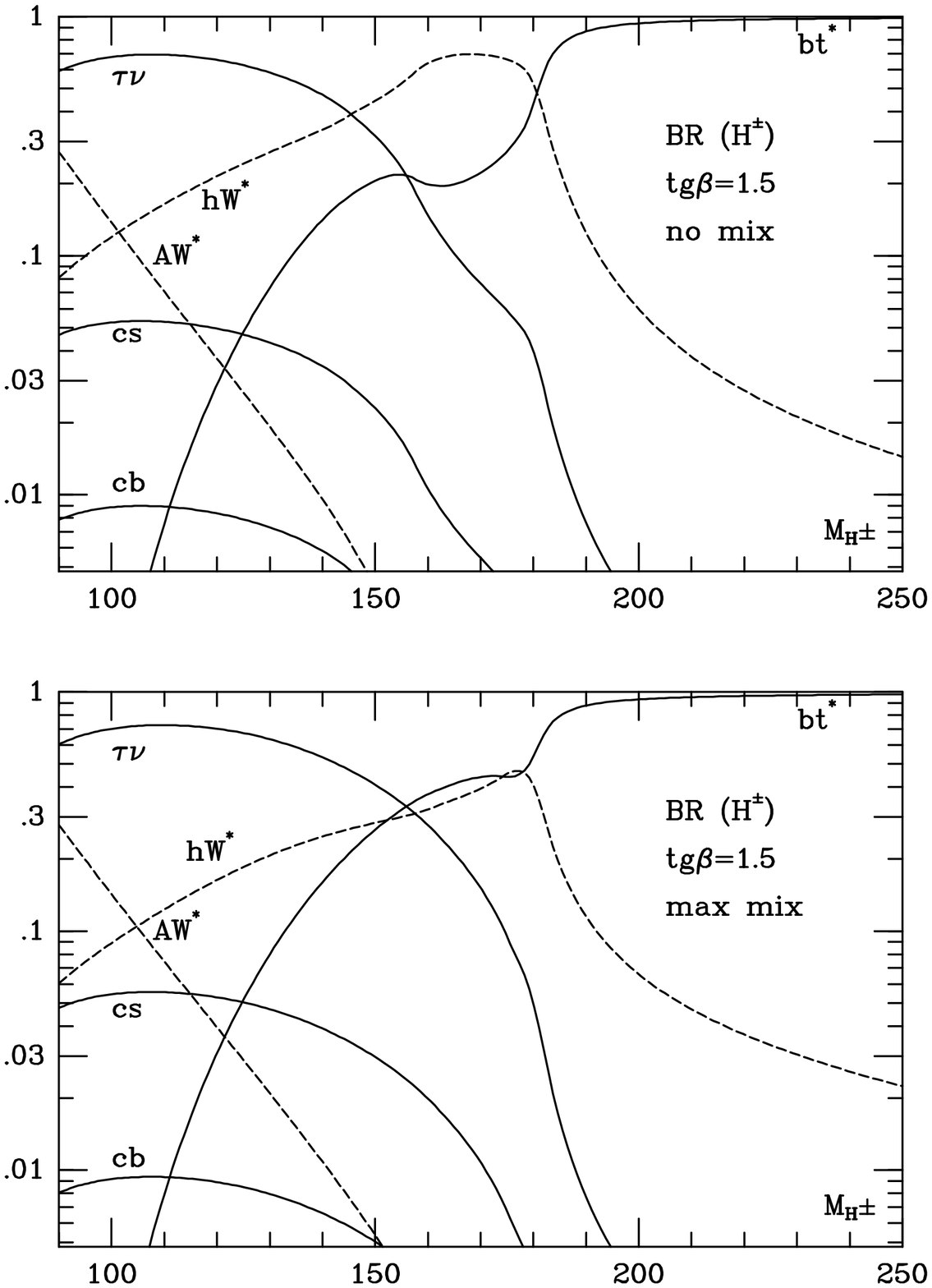,height=21.5cm,width=15cm}}
\vspace*{-3cm}
\centerline{\bf Fig.~6d}
\end{figure}

\newpage

\begin{figure}[htbp]
\vspace*{1cm}
\centerline{\psfig{figure=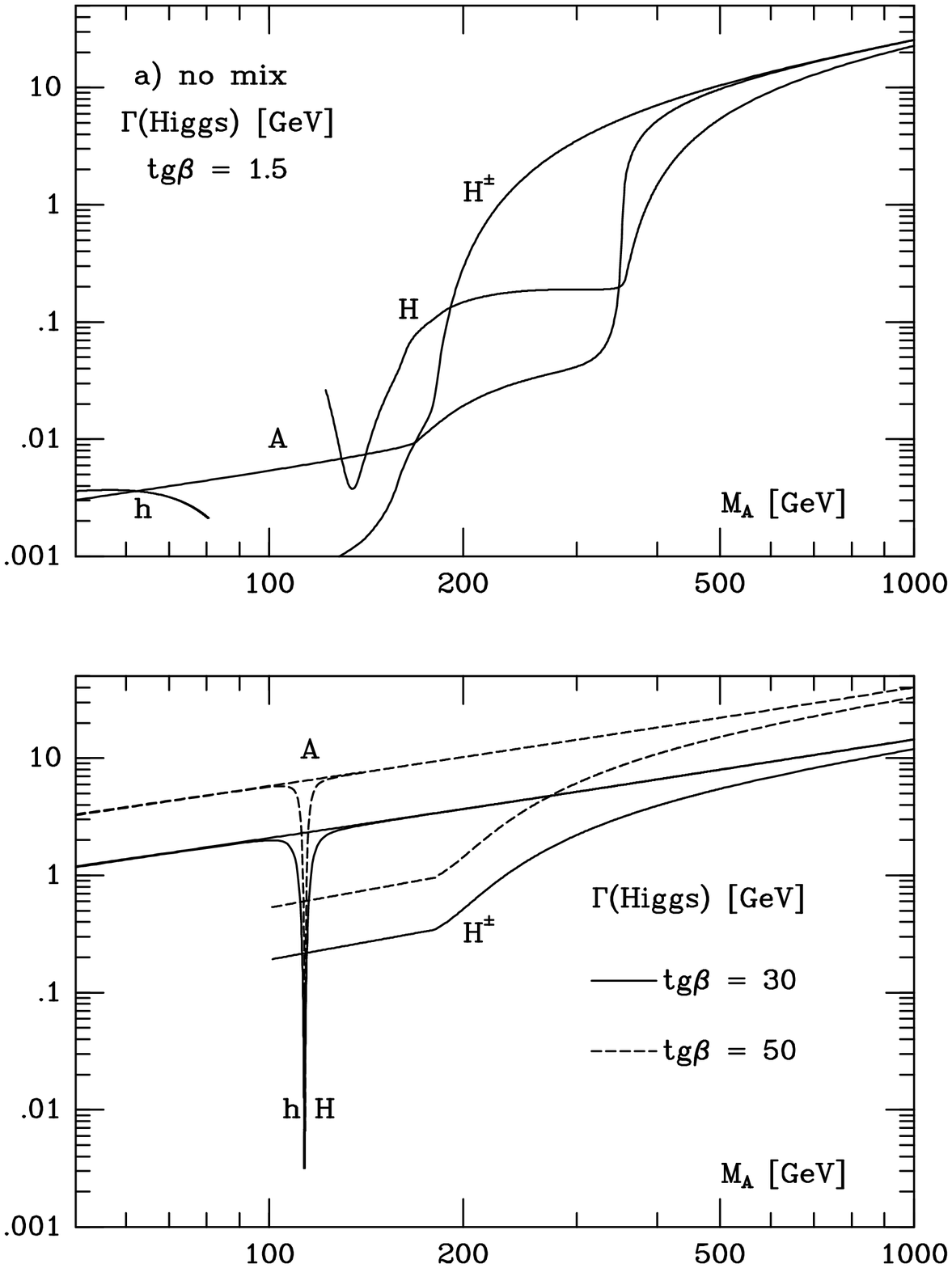,height=21.5cm,width=15cm}}
\vspace*{-3cm}
\centerline{\bf Fig.~7}
\end{figure}

\newpage

\begin{figure}[htbp]
\vspace*{1cm}
\centerline{\psfig{figure=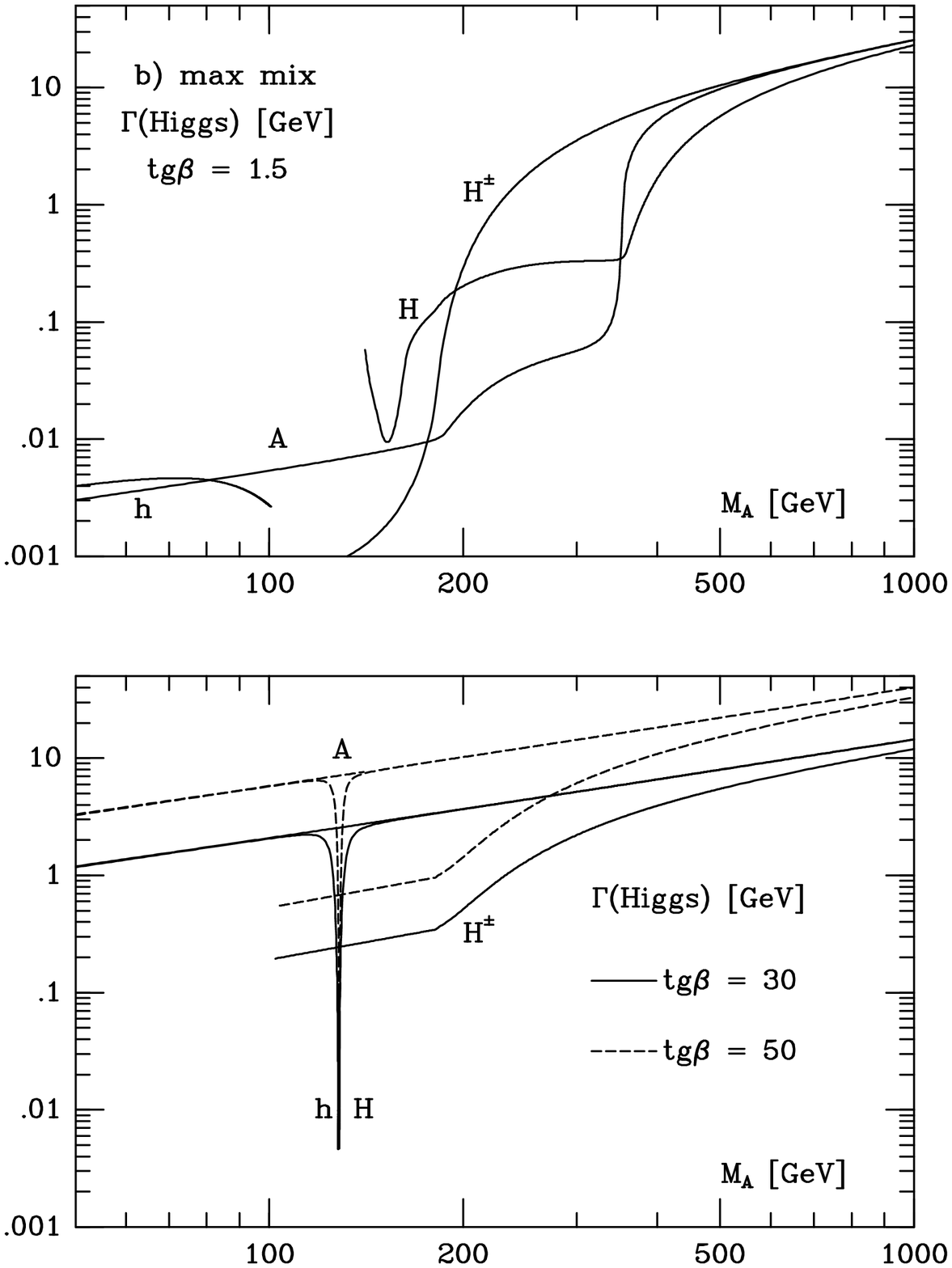,height=21.5cm,width=15cm}}
\vspace*{-3cm}
\centerline{\bf Fig.~7 (cont.)}
\end{figure}

\end{document}